\documentstyle[12pt,aasms4]{article}

\begin{document}

\title{Positron Escape from Type Ia Supernovae}

\author{P.A. Milne\footnote{NAS/NRC Resident Research Associate.}
\footnote{Much of this work was performed  at Clemson University.}}
\affil{Naval Research Laboratory, Code 7650, Washington, DC 20375\\}
\author{L.-S. The, M.D. Leising}
\affil{Department of Physics and Astronomy\\
    Clemson,S.C., 29634-1911}

\begin{abstract}
We generate bolometric
light curves for a variety of type Ia supernova models at late times,
simulating gamma-ray and positron transport for various
assumptions about
the magnetic field and
ionization of the ejecta.
These calculated light curve shapes are compared with
light curves of specific supernovae for which there have been
adequate late observations.
From these comparisons
we draw two conclusions: whether a suggested model is an acceptable
approximation of a particular event, and, given that it is, the magnetic
field characteristics and
degree of ionization that are most consistent with the observed light curve
shape. For the
ten SNe included in this study, five strongly suggest $^{56}$Co positron
escape as would be permitted by a weak or radially-combed magnetic
field. Of the remaining five SNe, none clearly show the upturned light
curve expected for positron trapping
in a strong, tangled magnetic field.
Chandrasekhar mass models can explain
normally, sub-, and super- luminous supernova light curves;
sub-Chandrasekhar mass models
have difficulties with sub- (and potentially normally) luminous
SNe. An estimate of the galactic positron production rate from type
Ia SNe is compared with gamma-ray
observations of Galactic 511 keV annihilation radiation.
Additionally, we emphasize the importance of correctly
treating the positron transport for calculations of spectra, or
any properties, of type Ia SNe at
late epochs ($\geq$ 200 d).

\end{abstract}


\section{INTRODUCTION}

Many years of  study of the light curves and spectra of type Ia 
supernovae have
revealed that a thermonuclear runaway 
induced by accretion on a white dwarf of $\sim$1 M$_{\sun}$ (Hoyle \&
Fowler 1960) produces the energy to eject the entire star at
$\sim$10,000 km s$^{-1}$
 and synthesizes $\sim$0.1-1.0 M$_{\sun}$ of radioactive
$^{56}$Ni.  Prior to the maximum luminosity, the
ejecta are opaque and both the explosion energy and the energy from the
$^{56}$Ni$\rightarrow^{56}$Co$\rightarrow^{56}$Fe decays are deposited in
and diffuse outward through the ejecta.  After the bolometric light curve
reaches its peak, the luminosity approaches the instantaneous decay power
(Arnett 1980; 1982) and the luminosity decline
follows the decay of $^{56}$Co, with $\gamma$-ray
scattering and absorption dominating energy
deposition.  Later, due to expansion, $\gamma$-ray optical depth decreases
and the bolometric light curve decline becomes steeper than the
$^{56}$Co decay.  Around $t \geq$ 200$^d$, almost all $\gamma$-ray
photons directly escape the ejecta and positrons from the decay of
$^{56}$Co provide the main source of energy
deposition into the ejecta. At this time, the light curve settles onto
another, lower, $^{56}$Co decay curve.
Beyond this time the light curve evolution depends on the details of
positron transport and energy loss.
Here we calculate energy deposition in type Ia supernovae (SNe Ia) 
for $t \geq$ 60$^{d}$, when the  photon diffusion time is short and the
bolometric luminosity is almost entirely due to gamma-ray and
positron energy loss processes.

There has been significant progress in
understanding the physics of the early type
Ia light curves. Detailed radiation transport 
and expansion opacity computations have reproduced observations of 
the bolometric and filter light curves (i.e., Pinto \& Eastman 1998; 
H\H{o}flich, M\H{u}ller, \& Khokhlov 1993; Woosley \& Weaver 1986).
It is clear that the light curve's rapid early evolution is due to 
the explosion of a low mass compact object with a short radiative
diffusion time (Arnett 1982; Pinto \& Eastman 1998).
The secondary IR-maximum is the result of a photospheric radius that
continues to expand after visual maximum.
The differences in pre-maximum rise-times among observed SNe
can be explained in model-dependent terms as due to the differences in a transition from 
a deflagration flame to a detonation.
The connections among peak luminosity, light curve width, and 
post-maximum slope with the ejecta mass, $^{56}$Ni location, 
kinetic energy of explosion, and opacity are explored in
an elegant paper by Pinto \& Eastman (1998). 
Comparisons between  early observed light curves and many various theoretical
models have been made extensively in recent years
( H\H{o}flich, Wheeler, \& Thielemann 1998; 
H\H{o}flich et al. 1996; H\H{o}flich \& Khokhlov 1996 (hereafter HK96)).
We take the results of these works as given and extend the comparison
of the models and observations to later times.

For a few bright SNe Ia in the past half-century, good 
light curves beyond $t \geq$ 200$^d$
have been obtained and studied.
The first work that utilized the information from 
late light curves was done by Colgate, Petschek, \& Kriese (1980).
They showed that the late photographic and B band 
light curves of SN 1937C \& SN 1972E can be
explained with positron kinetic energy deposition from $^{56}$Co decays,
as suggested by Arnett (1969).  Recently Colgate (1991, 1997) analyzed 
late light curves with positron transport to infer the ejecta
mass. They found that ejecta somewhat transparent to 
positrons was required
to fit observed light curves.  Cappellaro et al. (1997) (hereafter CAPP)
studied
the influence of the mass of ejecta and the total mass of
$^{56}$Ni  on SN Ia light curves, and concluded that the fraction of
positron energy deposition varies among the five SNe Ia they analyzed,
from complete transparency to a complete trapping of the positrons.
Ruiz-Lapuente \& Spruit (1998) (hereafter RLS) 
studied the effects of magnetic field
geometry and magnitude in exploding white dwarfs to explain
observed SNe Ia bolometric light curves.  They found that the  SN 1991bg
magnetic field was weak ($\leq$10$^3$G), while for SN 1972E a full trapping
of positrons by a tangled magnetic field in a Chandrasekhar mass model was
required.  

In addition to furthering our understanding of SNe Ia, we are motivated 
by the possibility that the Galactic diffuse 511 keV
emission might be due to the annihilation of positrons that escape from
SNe Ia, in addition to other possible sources such as core-collapse
SN radioactivity and compact objects.
The long lifetime of energetic 
positrons in the ISM (10$^3$-10$^7$ yrs)
would mean that positrons from many SNe Ia would appear as a diffuse
component of the 511 keV line emission.  Chan \& Lingenfelter (1993)
(hereafter CL)
calculated the number of surviving positrons from type Ia supernova
models for different magnetic field assumptions. Combining their
positron yields with supernova rate estimates,
they suggested that type Ia SNe could provide from an insignificant
to a dominant contribution to 
the observed diffuse Galactic annihilation radiation flux -- depending
on the magnetic field geometry. Average escape of a few per cent of $^{56}$Co
positrons are necessary  to contribute to the Galactic emission, i.e.,
$\geq$50\% of those emitted after one year would have
to escape. If the escapees take a significant fraction of their kinetic
energy with them, we might be able to see the power deficit in the
measured light curves. Thus we hope to use observed light curves to
determine the SNe Ia contribution to Galactic positrons and obtain
information about the magnetic fields in SNe Ia.
 
Because of the variety of observed SN Ia properties and numerous 
physical details that remain uncertain,
 there are many possible models allowed
(for a review see Nomoto et al. 1996). We do not try to make judgements
among them. We
calculate the late bolometric light curves of deflagration, delayed
detonation, pulsed-delayed detonation, He detonation, and
accretion-induced collapse (AIC) models for
comparison with observed light curves. For each observed SN we
consider the model(s) suggested by other authors to agree with
observations near maximum light. We begin by fitting each model
to the observations (to the inferred bolometric light curve when
available) in the interval 100--200 days, when gamma-ray energy
deposition is still important. We then follow each model to later
times for various choices of magnetic field configuration and
assumed ionization state of the ejecta, and compare the calculated
power deposited to the observed light curves.

In addition to the specific light curve consequences of positron 
transport we examine some more generic features of the late light
curves.
We show that the light curve decline rate between 60 and 200 days
after the explosion can be used to clarify total ejecta mass 
and 
expansion velocity of the supernova models. 
We find that the transition time 
between the gamma-ray and positron dominated phases 
is a useful diagnostic to measure the ejecta mass.
In addition, the slope of the light curve at $t\geq$200 days 
is a good indicator of the magnetic field configuration in the ejecta.  
As a baseline, we 
compare the calculated light curves from positron transport with the
light curve that results from the immediate, 
in-situ kinetic energy deposition assumption, which has been
 typically assumed in early light curve calculations.

The physics of energy deposition in type Ia SNe is discussed in section 2, 
followed by a discussion of the strength and geometry of the
magnetic field in the post-explosion ejecta. Our bolometric light
curves for both turbulent and radial field geometries are shown and
described in Section 3 for the various models used in our study.  
The issues confronting fitting model generated light curves to actual
observations are discussed in section 4. In
Section 5, we combine model predictions with SN data to fit a few of
the best observed SNe. We conclude with Section 6 in which we use
positron yields from our models to estimate the type Ia contribution to
the observations of 511 keV annihilation radiation in the Galaxy.


\section{Energy Deposition in SN Models}

The $^{56}$Ni $\rightarrow$ $^{56}$Co ($\tau \sim$ 8.8$^{d}$)
decay proceeds via electron capture (100\%) and produces
photons and neutrinos. The line photons have energies between
0.16 and 1.56 MeV, with a mean energy in photons of 1.75 MeV per
decay. The  $^{56}$Co $\rightarrow$ $^{56}$Fe ($\tau \sim$
111$^{d}$) decay proceeds 81\% of the time via electron capture
and 19\% via positron emission ($\beta^{+}$ decay). The line
photons have energies between 0.85 and 3.24 MeV, with  mean
energy in photons per decay of 3.61 MeV.  Segre (1977) derived  the
distribution of positron kinetic energy (which is shown in figure 
\ref{emit}). The mean positron kinetic energy is 0.63 MeV,
so when multiplied by the branching ratio 0.19, the mean
positron energy per decay is 0.12 MeV. The ratio of the mean
photon energy to the mean positron kinetic energy is then $\sim$ 30. The photon
luminosity will dominate until the deposition fraction for
positrons (f$_{e^{+}}$) is $\sim$ 30 times larger than the
deposition fraction for photons (f$_{\gamma}$). The energy
deposition from both gammas and positrons  depend upon two
factors; how much material must be traversed enroute to the
surface, and the interaction cross-sections.
The transport of gamma rays and positrons will  be
developed separately.

\subsection{Gamma-Ray Transport}

The gamma-line photons from the decays of 
$^{56}$Ni$\rightarrow ^{56}$Co$\rightarrow ^{56}$Fe are mostly
Compton scattered to lower energy during the early phases of
the Type Ia event. These photons then escape as X-ray continuum
or are absorbed by material in the ejecta via the
photo-electric effect.  Due to the supernova expansion, the
gamma-line optical depth decreases, becoming low enough at
late-times  that most gamma-line photons escape directly.  In
calculating the energy deposition and the energy escape
fraction of the photon energy created by the radioactive
decays, we simulate the scatter adopting the prescription of
Podznyakov, Sobol, \& Sunyaev (1983). A detailed description of
the Monte Carlo algorithm and  its application in calculating
the spectra and bolometric light curves of  SN1987A and type Ia
supernovae have been presented by  The, Burrows, \& Bussard
(1990) and Burrows \& The (1990).

\begin{figure}
\plotone{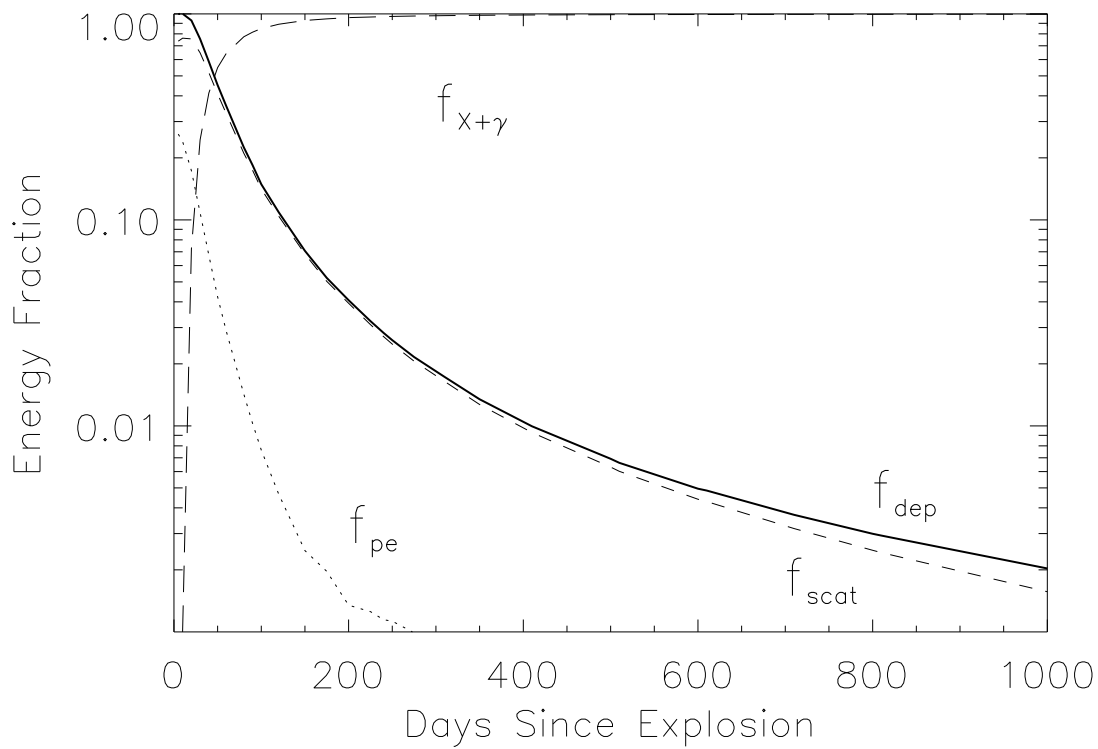}
\caption{The outcomes of gamma and X-ray photon interactions in the
ejecta. The fraction of the decay energy that is deposited via
scattering is labeled $f_{scat}$, the fraction deposited via
photoelectric absorption is labeled $f_{pe}$. The total fraction
deposited is the sum of these two fractions and is labeled 
$f_{dep}$. Scattering dominates the energy deposition until
$f_{dep}$ lowers to 0.03, at which time positron kinetic energy
deposition dominates.
 \label{fracE}}
\end{figure}

Figure \ref{fracE} shows the fraction of the total decay energy
that escapes directly as gamma-lines and  emerges as X-ray
continuum (f$_{\gamma +X}$) as  well as the fractions deposited
by scattering (f$_{scat}$) and  photoelectric absorption
(f$_{PE}$) as functions of time for model W7 (Nomoto,
Thielemann \& Yokoi 1984), which we use for illustration here.
The solid curve labeled with f$_{dep}$ (= f$_{scat}$ +
f$_{PE}$) is the total gamma-ray energy deposition fraction.
Scattering dominates over photoelectric absorption at all
times, the photoelectric absorption becoming relatively less
important as there become fewer multiply scattered photons at
late times. The overall photon deposition fractionscales roughly as  
t$^{-2}$, reflecting the time dependence of the column depth to
the surface from any location in the ejecta.

\subsection{Positron Transport}

\begin{figure}
\plotone{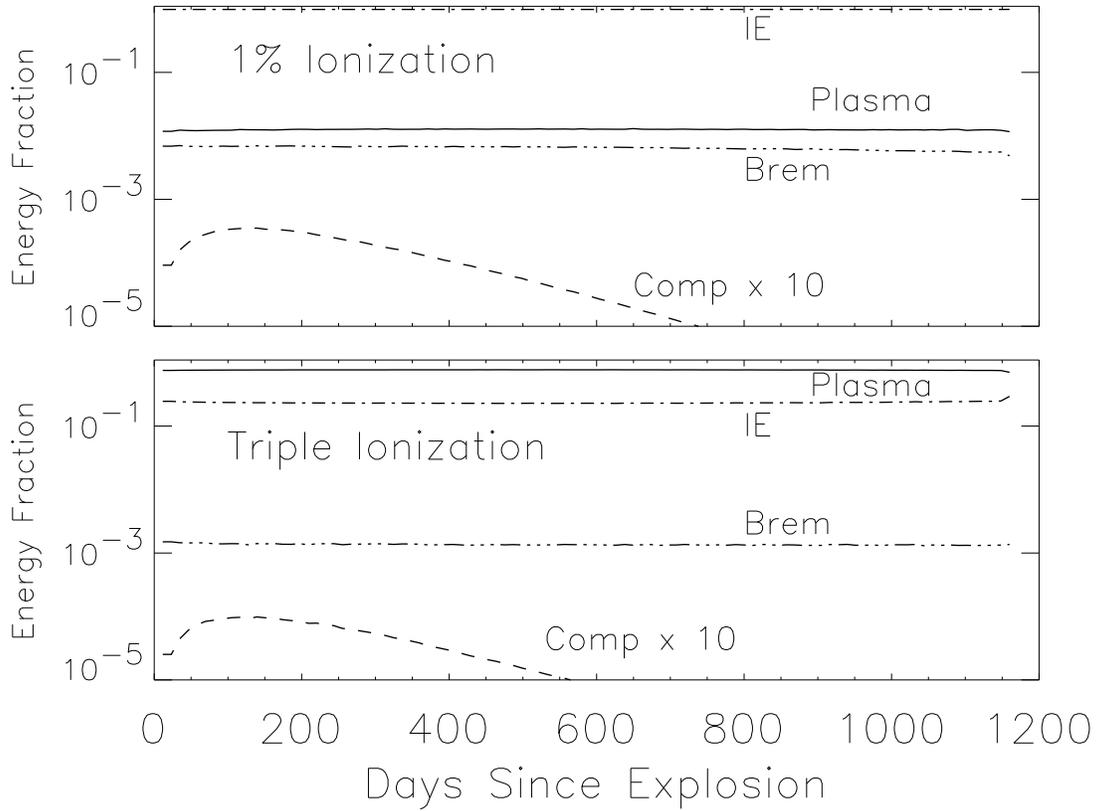}
\caption{The relative contributions of various modes of positron kinetic
energy loss. The upper plot shows the energy losses for 
1\% ionized ejecta, demonstrating that  ionization and excitation of bound
electrons dominates. The lower plot shows the energy losses for triply
ionized ejecta. For higher 
levels of ionization, energy losses to the plasma equal and then
dominate over the energy losses to bound electrons. \label{eloss}}
\end{figure}

The cross sections for the various modes of positron
annihilation are strongly energy-dependent and favor low
energies. Energetic positrons  will move through the ejecta
until slowed to thermal velocities and then quickly
annihilate.\footnote{The  lifetime of positrons once
thermalized is small compared to the slowing time and to the
decay timescale, so for this study we set it equal to zero.}
Five energy-loss mechanisms were included: ionization and
excitation of bound electrons, synchrotron emission,
bremsstrahlung emission, inverse Compton scattering, and plasma
losses. The first four are described in detail in CL.  We
include these processes in a calculation of model W7, first
assuming a uniformly 1\% ionized medium (i.e., 0.01 free
electrons per nucleus) and then a uniformly triply ionized
medium. The results, shown in figure~\ref{eloss}, support CL's
conclusion that for low ionization, the interactions with bound
electrons dominate the energy loss. For   higher ionization, we
find that plasma energy losses must be included. Regardless of
the ionization, processes other than  ionization and excitation
and plasma losses can be ignored.

The ionization and excitation energy loss rate used is (Heitler
1954; Blumenthal $\&$ Gould 1970; Gould 1972; Berger $\&$
Seltzer 1954),

\begin{displaymath}
    \left( \frac{dE}{d \xi}
    \right)_{ie} = - \frac{4 \pi r_{o}^{2} m_{e} c^{2} Z_{B}}
{\beta ^{2} A
m_{n}} \left[ ln \left( \frac{ \sqrt{\gamma -1} \gamma \beta}{\frac{
\overline{I}}
{m_{e}c^{2}}}    \right)
+ \frac{1}{2} ln2
       \right)
    \end{displaymath}

\begin{equation}
       \left.
 - \frac{ \beta^{2}}{12} \left(
\frac{23}{2} + \frac{7}{\gamma +1} + \frac{5}{ (\gamma + 1)^{2}} +
\frac{2}{ (\gamma +1)^{3}}                              \right)
       \right],
   \end{equation} 

\noindent where E is the positron's kinetic energy, r$_{o}$ is the
classical electron radius, and m$_{e}$ and m$_{n}$ are the
electron and atomic mass unit masses respectively. Z$_{B}$ and
A are the effective nuclear charge and atomic mass of the
ejecta. $\overline{I}$ is the effective ionization potential
and is approximated by Segre (1977) and by Roy $\&$ Reed(1968)
as,

\begin{equation}
    \overline{I} = 9.1 Z \left( 1 + \frac{1.9}{Z^{\frac{2}{3}}}
              \right) eV .
\end{equation}

The plasma energy loss was described by Axelrod (1980). The
formula is the same as  ionization and excitation, except 
$\hbar \omega_{p}$ is inserted in the calculation of the 
maximum impact parameter (b$_{max}$) rather than the   mean
ionization potential ($\overline{I}$), and $\chi_{e}$, the 
number of free electrons per nucleus (hereafter ionization 
fraction) is used rather than Z$_{B}$.  Thus, (Z$_{B}$ +
$\chi_{e}$)$\cdot$n = Z$\cdot$n is  the total electron density.
Ignoring small differences in the relativistic  corrections,
when the two energy losses are summed,

\begin{displaymath}
\left( \frac{dE}{d \xi}
    \right)_{Total} = \left( \frac{dE}{d \xi}
    \right)_{ie} + \left( \frac{dE}{d \xi}
    \right)_{Plasma}
\end{displaymath}
\begin{equation}
 =  - \frac{4 \pi r_{o}^{2} m_{e} c^{2} Z P(E)}
{\beta ^{2} A m_{n}} 
\left[Z ln \left( \sqrt{\gamma -1} \gamma \beta m_{e}c^{2}
 \right) -Z_{B} ln \overline{I} -\chi_{e} ln (\hbar \omega_{P}) \right],
\end{equation}

\noindent where P(E) is the relativistic correction. There is
an ionization fraction dependence due to the inequality between
$ln \overline{I}$ and $ln (\hbar \omega_{P})$. An ionized
medium is more efficient at slowing positrons than is a neutral
medium. The ionization fraction must be known  as a function of
mass, radius, and time  to determine the energy deposition
exactly, but  is not currently well understood. Pinto \&
Eastman (1996), Liu, Jeffery \& Schultz (1997), and Fransson 
\& Houck (1996) arrived at different ionization structures for
similar models. Unable to improve this situation, we choose to
calculate a range of  ionization fractions to bracket the
possibilities.  We consider extreme values of the ionization
fraction, 0.01 and 3.0, because the actual values are almost
certainly between these during the times we consider.

The positron transport was done with a 1D Monte Carlo code. SN
models were reduced to 75-150 zones and up to 34 elements.
Positrons were emitted at equally spaced time  intervals from
each zone (volume-weighted, random locations within the zone),
weighted according to the $^{56}$Ni(zone,t). Positrons were
followed in steps of equal column depth; the time, energy,
radial distance, zone, pitch angle, and annihilation
probability were re-evaluated at each step. 

\begin{figure}
\plotone{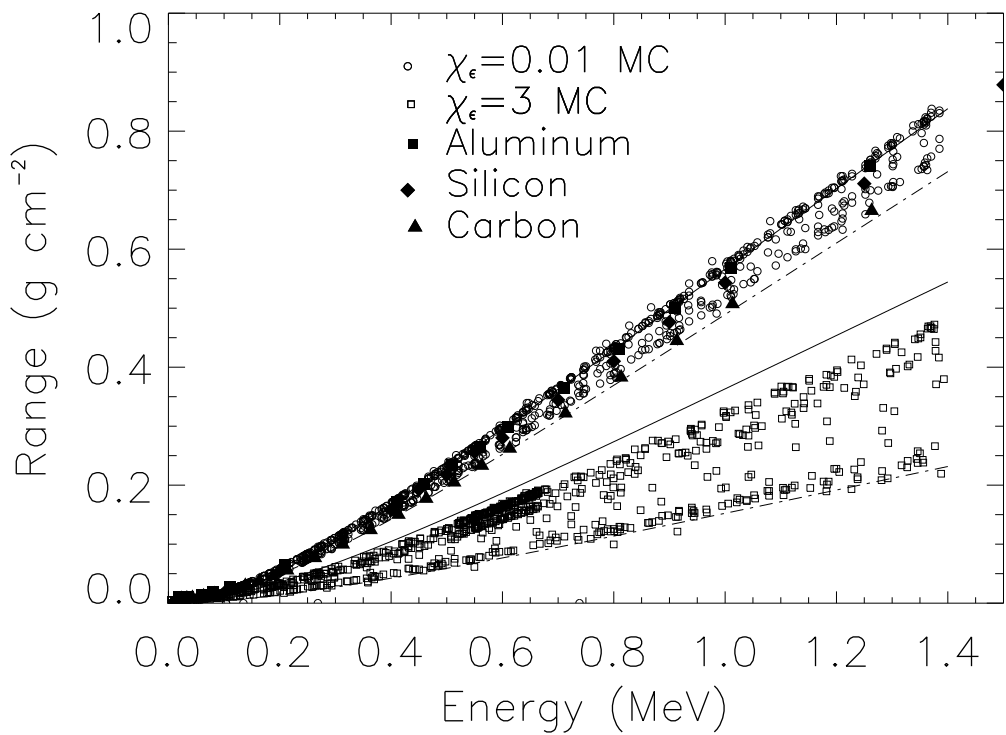}
\caption{The ranges of non-thermal
 positrons from MC sampling, assuming 1\% (open circles) 
and triply (open boxes) ionized  
conditions compared with laboratory results in solid media.
 It is evident that highly ionized gas is more effective at
stopping positrons than is a solid. The solid lines are calculated 
ranges using equation 3 for each level of ionization 
and assuming the ejecta is 
pure Fe, the dot-dashed lines are the ranges in pure C.
The laboratory measured ranges for solid 
Aluminum, Silicon and Carbon are taken from ICRU Report \#37
(1984). The neutral ejecta stops positrons approximately as
does a solid medium. \label{range}}
\end{figure}

The range of positrons  in various solid  media has been
measured in the laboratory (ICRU Report \#37 (1984)). Figure
\ref{range} compares the laboratory results with the  ranges of
individual positrons through SN ejecta, combining the data from
a number of models. The model results for low ionization are in
good agreement with the laboratory results,  whereas the higher
ionization leads to the medium being 2-3 times more efficient
at stopping positrons.  The spread of ranges for the triply
ionized ejecta reflects the variation of the ejecta's
composition. A positron will have a longer range in triply
ionized iron than in triply ionized carbon because of the
$\frac{\chi_{e}}{A}$ dependence of the more efficient energy
loss to the plasma. It is unlikely that the zones rich in C, O,
Si and the other intermediate elements will maintain as high a
level of ionization as the Fe zones (where the decays occur),
so the uniformly triply-ionized approximation is a lower limit
for positron escape. The 1\% ionized ranges have less scatter
because the ionization-excitation range depends upon
$\frac{Z_{B}}{A}$, and with $Z_{B} \approx Z$, $\frac{Z_{B}}{A}
\approx$ 0.5 in both Fe-rich zones and C-rich zones. 

The range determines the escape fraction because  when the
column depth to the surface at a given radius is less than the
stopping power for a given emission  energy, a positron of that
energy can escape (after it deposits a fraction of its energy 
in the ejecta).  A number of groups have transported positrons
with the same treatment used with photons, assigning a
$\kappa_{e+}$, an ``effective  positron opacity", incorrectly
giving them an exponential distribution of ranges. We do not
use $\kappa_{e+}$, but we can estimate its value for comparison
from our results.  Defining that $\kappa_{e^{+}} \equiv$ the
mean of the inverse of the range expressed in units of inverse
column depth (cm$^{2}$ g$^{-1}$);  for $\chi_{e}$ = 0.01,
$\kappa_{e^{+}}$ = 4; while for $\chi_{e}$ = 3, 
$\kappa_{e^{+}}$ = 14.  Colgate transported positrons with
$\kappa_{e^{+}}$ = 10 cm$^{2}$ g$^{-1}$, as did Ruiz-Lapuente.
Axelrod argued for $\kappa_{e^{+}}$ = 7 cm$^{2}$ g$^{-1}$,
which was the nominal value in most fits shown by CAPP.

\subsection{Magnetic Field Considerations}

The efficiency of matter at slowing positrons is one factor in
determining the positron escape, the quantity of matter
traversed enroute to the surface is another. The effects of
the total mass and the nickel distribution will be discussed in
section 4, but these are secondary to the effects of the
magnetic field (CL), which will force positrons to follow curved
paths. The progenitor white dwarfs have been observed to have
initial field strengths of 10$^{5}$ -10$^{9}$ Gauss (Leibundgut
1995). The field is assumed to diffuse on a long time-scale
relative to the positron lifetimes, so the flux is treated as
frozen-in. Thus the expansion causes the  field strength
to evolve according to the relation,
 B(r) = B(r$_{o}$) r$_{o}^{2}$ r$^{-2}$.
If the resulting field is strong enough to make the positron
gyroradius smaller than the ejecta at a given time, the
geometry of the field must be considered. Little is known about
SN Ia magnetic fields, so we consider three scenarios; radial
magnetic field, disordered field, no field. Thus we also
bracket the  extreme magnetic field configurations, and late
observations of SN Ia luminosities are potentially probes of
the field characteristics.

A radial field is an approximation of the effect of the rapid,
homologous  expansion of the ejecta stretching out the
arbitrarily oriented field lines as it changes scale by a
factor 10$^{7}$ as it grows. Positrons spiral inward or outward
on these field lines, changing pitch angle due to mirroring and
beaming, as described by CL. These bend trajectories  radially
outward, so  this ``radial scenario'' is a favorable one for
positron escape.

A turbulent, confining field is presumably  adjusted or
generated by expansion dynamics (RL \& Spruit 1997),  with the
limiting case of a positron having no net radial motion (in
mass coordinates)  as it meanders near the location of its
emission zone for all times. This is called the ``trapping
scenario". There is some survival of positrons in place, if
their slowing times exceed the age of the supernova, but by
definition there is no escape.

The third scenario ignores magnetic fields, so the positrons
travel straight-line trajectories. This is referred to as the
``free scenario".

\section{Bolometric Light Curves}

Combining the results of the gamma-ray and positron transport
calculations, we obtain the rate of energy deposition throughout
the SN ejecta.  This  energy is then mainly contained in
suprathermal free electrons that slow and recombine into atoms,
which then deexcite. As discussed for positrons above, once the
electrons reach energies of a few keV their slowing times, and
the duration of the subsequent processes are short. At late times
all the optical photons generated escape immediately without
interaction. So, for the time of interest here, the major
difference between the decay rate of the radioactivity and the
supernova (optical) luminosity  is the propagation times of the
gamma rays and positrons, and we therefore treat the calculated
power deposited and bolometric luminosity as being interchangeable. A
typical calculated bolometric light curve (power deposition) 
is shown in figure
\ref{bol}. 

\subsection{Time Evolution 
\protect\footnote{Unless otherwise stated, times quoted refer to time since
explosion}}

The initial climb to peak luminosity is generated by  energy
deposition from the thermonuclear burning and from the  
$^{56}$Ni decays, followed by the diffusion of light to the
photosphere. The width of the peak is governed by the diffusion
time from the $^{56}$Ni to the photosphere.  Observed SNe have
shown variations of peak width, which should arise naturally out
of any successful models.
The factors that
determine the light curve peak width are currently debated. 
H\H{o}flich (H\H{o}flich \& Khokhlov 1996) asserts
that the ejecta mass, the $^{56}$Ni  distribution, and the $^{56}$Ni 
mass all influence the peak width. He fits narrow peaks with
models which underproduce $^{56}$Ni  (relative to W7). Pinto \&
Eastman (1996) asserts that the peak width only depends upon the
first two.  Their radiation transport
calculation showed that the energy
deposition from the $^{56}$Ni  adjusted the structure of the ejecta
to make the peak width independent of the $^{56}$Ni 
production. They explain narrow peaks with low-mass models. 

Because we do not do the optical radiation transfer, we can not 
apply our calculations to this early epoch. 
We simply
take models shown by other authors to fit individual supernova 
early light curves and spectra and calculate their subsequent
bolometric light curves to be compared with late observations.
By 60 days after the explosion, the photosphere has receded to
the SN center; after this time our calculations should
trace the bolometric light curve.
\footnote{SN 1991bg may be the explosion of a
0.6 M$_{\odot}$ WD. If so, bolometric curves may fit the data as
early as 30 days. The choice of 60$^{d}$ as the time to initiate 
light curve fitting follows from Leibundgut \& Pinto 1992.}

\begin{figure}
\plotone{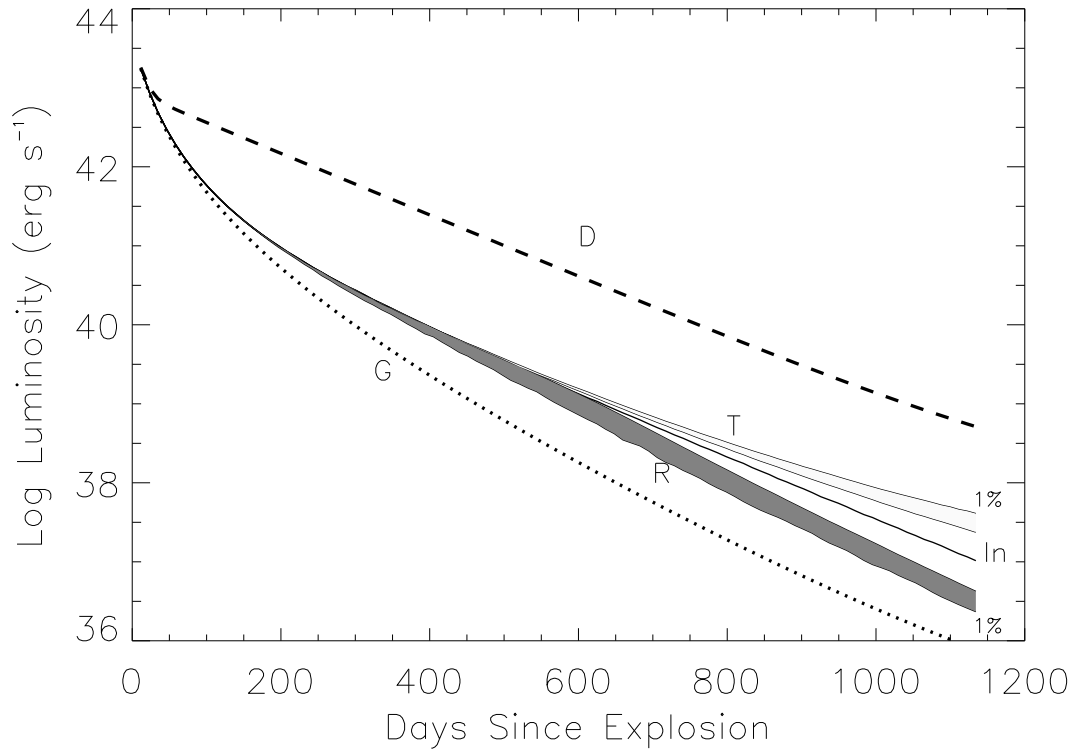}
\caption{A model generated bolometric light curve for the model W7. The
dashed line (D) assumes instantaneous deposition of all decay energy. The
dotted line (G) uses the results of the gamma energy deposition 
only and assumes
no deposition of positron energy. Between these two boundaries are the
results of the gamma energy deposition coupled with instantaneous
positron deposition (thick line, In) and the range of curves for a  radial
field geometry (dark shading, R) and for a trapping geometry light (
shading, T) as the electron ionization 
fraction varies from 0.01 $\leq \chi_{e} \leq$
3. \label{bol} }
\end{figure}

\begin{figure}
\plotone{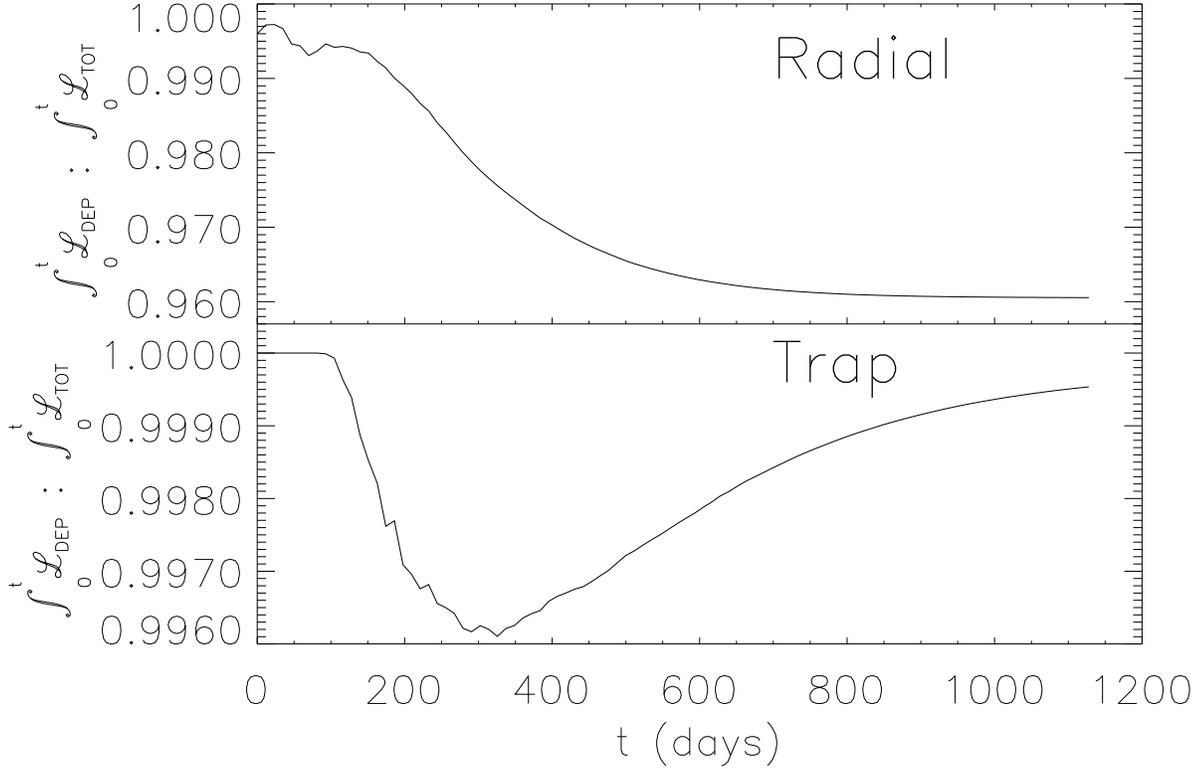}
\caption{The 
time-integrated luminosities for the radial and trapping field
geometries. The radial field permits energy to leave the system at times
late enough for positron transport, so only about 96\% of the 
accumulated energy is
deposited by 1000$^{d}$. The trapping field does not permit energy to
leave, but energy is stored as the positrons develop appreciable
lifetimes. The energy storage is at most 0.4\% of the total energy,
 and at later times most of that energy
is deposited, leading to 99.9\% of the energy being deposited by
1000$^{d}$. This late deposition leads to the flattening of the light
curve for the trapping field. \label{intl} }
\end{figure}

The light curve decline after 60$^{d}$ is governed by the
$^{56}$Co decay and the falling gamma-ray optical depth. Thus the
light curve shape is a diagnostic of the mass overlying the
Ni-rich zones  and the velocity structure.  The decline of the
gamma deposition fraction (f$_{\gamma}$) makes the light curve
steeper than the decay. Faster  models  reach the transition to
positron dominated power,  when f$_{\gamma}$ = 0.03, earlier. The 
time of this transition varies among models, occurring  at
130$^{d}$ -520$^{d}$. The ejecta is still too dense at these
times for the positrons to have appreciable lifetimes, so the
light curves flatten toward the $^{56}$Co decay line (RLS).
During this epoch, the various field geometry scenarios produce
identical results. Further expansion of the ejecta leads to
appreciable positron lifetimes, and in the radial and free
scenarios, escape. Then the light curves begin to separate. For
the radial and free scenarios, positron escape leads to kinetic
energy loss from the system and a drop in the light curve. For
the trapping scenario, a finite positron lifetime combined with
the exponential decline in the number of newly created positrons
leads to a shallow dip followed by a late flattening of the light
curve, as positron kinetic energy is stored and delivered a
(lengthening) positron slowing time after emission. Figure
\ref{intl} contrasts the integrated luminosities for the trapping
and radial scenarios as compared to the instantaneous deposition
approximation.  The delayed luminosity is seen in figure
\ref{bol} as the trapping curve is brighter than the
instantaneous deposition curve at late-times. For the radial
scenario, positrons (and thus kinetic energy) can leave the
system, leading to a much larger deviation from the instantaneous
deposition. As this energy is increasingly leaving the system at
late-times, the radial light curve in figure \ref{bol} is dimmer
than either the instantaneous deposition curve or the trapping
curve. As the radial and trapping curves diverge, the separation
becomes great enough to be measurable, in principle.

\subsection{Radial Magnetic Field vs No Field}

It turns out that there is surprisingly little difference in the
mean path traversed to the surface for these two cases. Assuming
isotropic emission of positrons, the mean distance to the surface
from a given point, for positrons not near the center of the
ejecta, is substantially larger than the radial distance to the
surface because of the large solid angle perpendicular to the
radius. In the radial field case, mirroring and beaming turn
positron trajectories outward, but the extra path due to the
spiral around the field adds similar distance as in the no-field
case. The net energy deposition and positron escape for the
models we consider are almost indistinguishable for these two
cases. This result was anticipated by Colgate (1996), and can be
approximately demonstrated analytically. We display radial field
calculations, but remind the reader that for our spherically
symmetric models the conclusions apply equally to field-free
scenarios. For a few of the models, there were slight deviations
at late-times (t $\geq$ 800$^{d}$, the field-free curves
remaining steeper than the radial curves). We show field-free
light curves only when the separation between  these these two
cases is potentially detectable.

\subsection{Energy Deposition at Very Late Times (~1000$^{d}$)}

\begin{figure}
\plotone{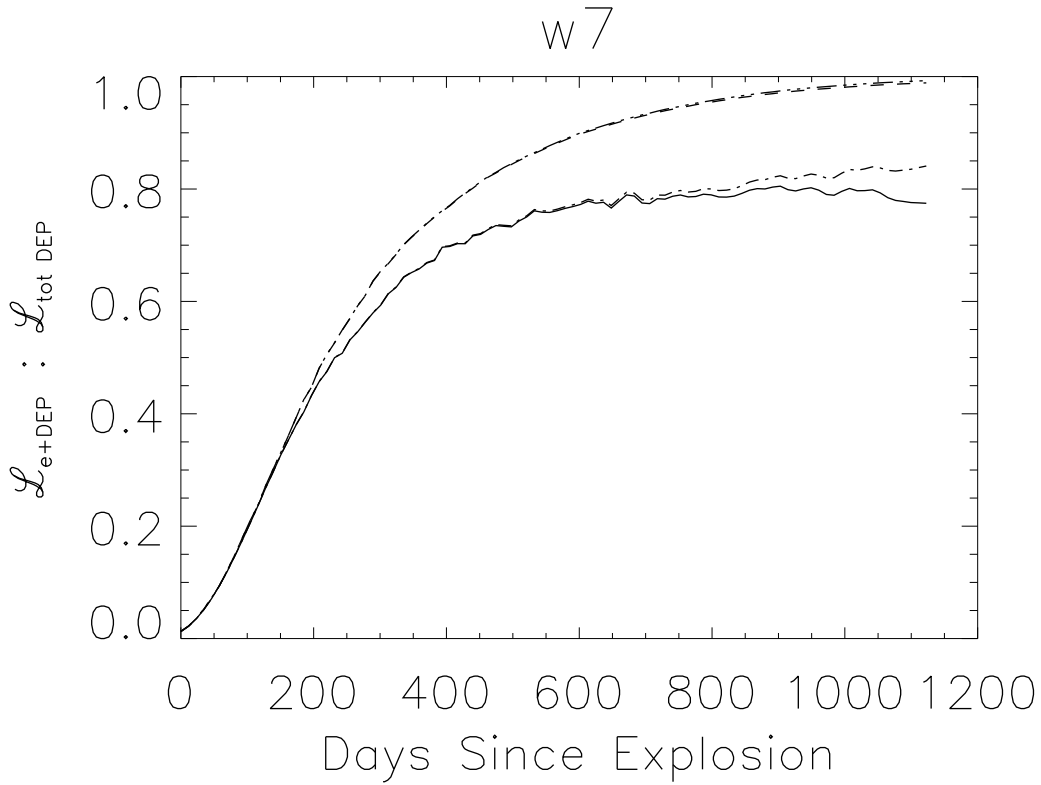}
\caption{The fraction of the total energy deposition rate that is
contributed by the slowing of positrons for the two field geometries.
The solid curve is for a radial field with $^{57}$Co photons, 
the dot-dashed curve is without $^{57}$Co 
photons. There is a period of time during which the slowing of positrons 
dominates the energy deposition. For W7, this period begins after
$\sim$250$^{d}$, when the gamma deposition fraction falls below 0.03. 
For the radial field, at later times positron escape lowers
the relative contribution of positron energy deposition. For the
trapping field (plotted with long and short dashed lines),
 positrons dominate out past 1000$^{d}$, regardless of
the contributions of $^{57}$Co photons. \label{pfrac} }
\end{figure}

The effects of positron escape on the SN light curve  are
moderated somewhat by gamma-ray energy deposition.  
Longer-lived radioactivities that were overwhelmed at earlier
epochs become important at late times. Two such radionuclei are
$^{57}$Ni and $^{44}$Ti.  The decay of $^{57}$Ni proceeds with
$\tau_{Ni57}$ = 52$^{h}$ and $\tau_{Co57}$ = 392$^{d}$, thus at
500$^{d}$, 28\% of the $^{57}$Co has yet to decay. The $^{57}$Co
decay energy is only 1/10 that of the $^{56}$Co. For W7 the
energy available from $^{57}$Co equals that available from
$^{56}$Co at 1000$^{d}$. Other models have similar
$^{57}$Ni/$^{56}$Ni ratios. $^{44}$Ti is an even longer-liver
radionucleus, with a mean-life recently estimated to be 85$^{y}
\pm$ 1$^{y}$ (Ahmad et al. 1997). The 1.3 MeV decay energy is
substantial, but the slow decay rate delays the cross-over to
$^{44}$Ti dominated deposition until 2500$^{d}$, when no SN Ia
has been detected. Five models had the $^{57}$Co and $^{44}$Ti 
masses included. For the other models, M$_{Co57}$ = 0.041 M$_{Co56}$ 
and M$_{Ti44}$ = 1.5 x 10$^{-5}$ M$_{Co56}$, to match W7. 
Figure \ref{pfrac} shows the fraction of the
luminosity that is due to the deposition of positron KE.  The
positron deposition is dominant from 300$^{d}$ -800$^{d}$ for the
radial field geometry.  The effects of including longer-lived
radioactivities is shown by the splitting of the curves at late
times.

\subsection{Additional Potential Sources and Effects}
There are a number of other potential sources of luminosity,
 both intrinsic and extrinsic.
One extrinsic source is the ``light echo'', bright peak light scattered
by dust back into the line of sight after light travel delays. SN 1991T
was dominated by a light echo by day 600 (Schmidt et al. 1994), 
as discussed in section 4.
Another 
extrinsic source is the interaction of the ejecta with the
surrounding medium. This interaction will eventually be important, but
should be identified by distinctive spectral and temporal characteristics.

The SN model light curves presented in this work are based on the 
assumption that the deposited power is instantly radiated
in the UVOIR bands, during the time 1--2 years. 
Furthermore, during this epoch,
we usually have only the V and/or B band observations,
instead of the bolometric luminosity, with which to compare
models. Fitting model light
curves to single or dual bands is susceptible to intrinsic
spectral evolution effects.
As the secondary electrons' lifetimes increase and the collisional 
de-excitation rates fall, a delay develops between the positron energy
deposition and emission of UVOIR light, an effect called
``freeze-out" by Fransson (1996). If the Fe I and II states form in 
abundance without photoionization or charge exchange destruction, 
then [FeII]$\lambda$25.99$\mu$, $\lambda$35.35$\mu$, and 
[FeI]$\lambda$24.05$\mu$ fine structure emission lines are produced. 
These lines are beyond the UBVRI bands and are undetected. This effect
is referred to as an infrared catastrophe (IRC) by Axelrod (1980). 
Both freeze-out and IRC phenomena must be considered and are 
discussed in section 5. It is clear that at very late times, many
complicating effects, including new sources of luminosity and
sinks outside the normally observed bands can be important.
Therefore we confine our conclusions to the times when $^{56}$Co 
decay dominates the power input and the B and V bands track well
the bolometric luminosity.

\subsection{SN Models}

We considered twenty-one models, which span the range
of ejecta mass, $^{56}$Ni  mass, and kinetic energy. 
One deflagration (W7) (Nomoto, Thielemann \& Yokoi 1984),
 a normally luminous 
helium detonation (HED8) (HK96),
 two subluminous helium detonations (WD065, HED6) (Ruiz-Lapuente et al.
1993; HK96),
two  superluminous helium detonations (HECD, HED9) (Kumagai 1997; HK96),
one accretion-induced collapse (ONeMg) (Nomoto 1996), 
 eight delayed or late detonations
(DD4, M36, M35, M39, DD2O2C, DD23c, W7DN, W7DT)  (Woosley 1991; H\H{o}flich
1995; H\H{o}flich, Wheeler \& Thielemann 1998; Yamaoka 1992), 
three pulsed, delayed detonations (PDD3, PDD54, PDD1b) (H\H{o}flich,
Khokhlov \& Wheeler 1995), 
and two mergers (DET2, DET2ENV6) 
were included. Their characteristics are shown in Table 1. 

\begin{figure}
\plotone{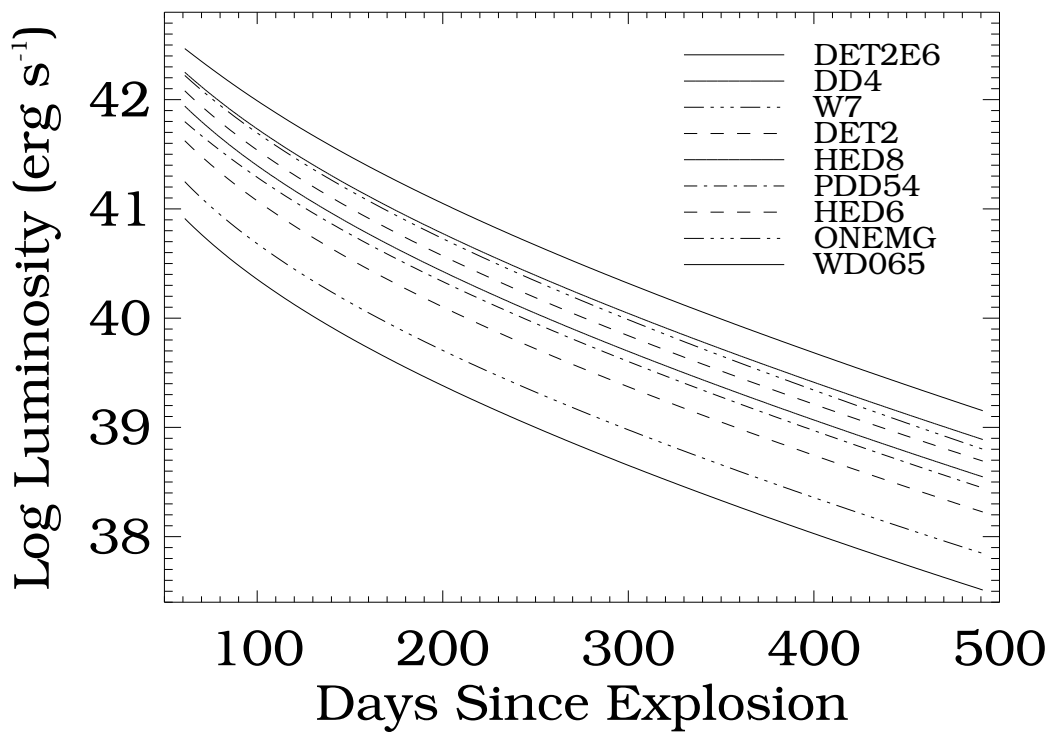}
\caption{The rate of energy deposition from the interactions of
$^{56}$Ni and $^{56}$Co decay photons with the ejecta. The overall
luminosity measures the amount of nickel in the model, the decline from
60$^{d}$ -150$^{d}$ is a measure of the amount of ejecta overlying the
nickel and its expansion velocity. 
The curves do not start at the peak because we have not treated
photon diffusion, an effect important before 60$^{d}$. Note the
similarity of curve shapes for all models. \label{gamma} }
\end{figure}

\begin{figure}
\plotone{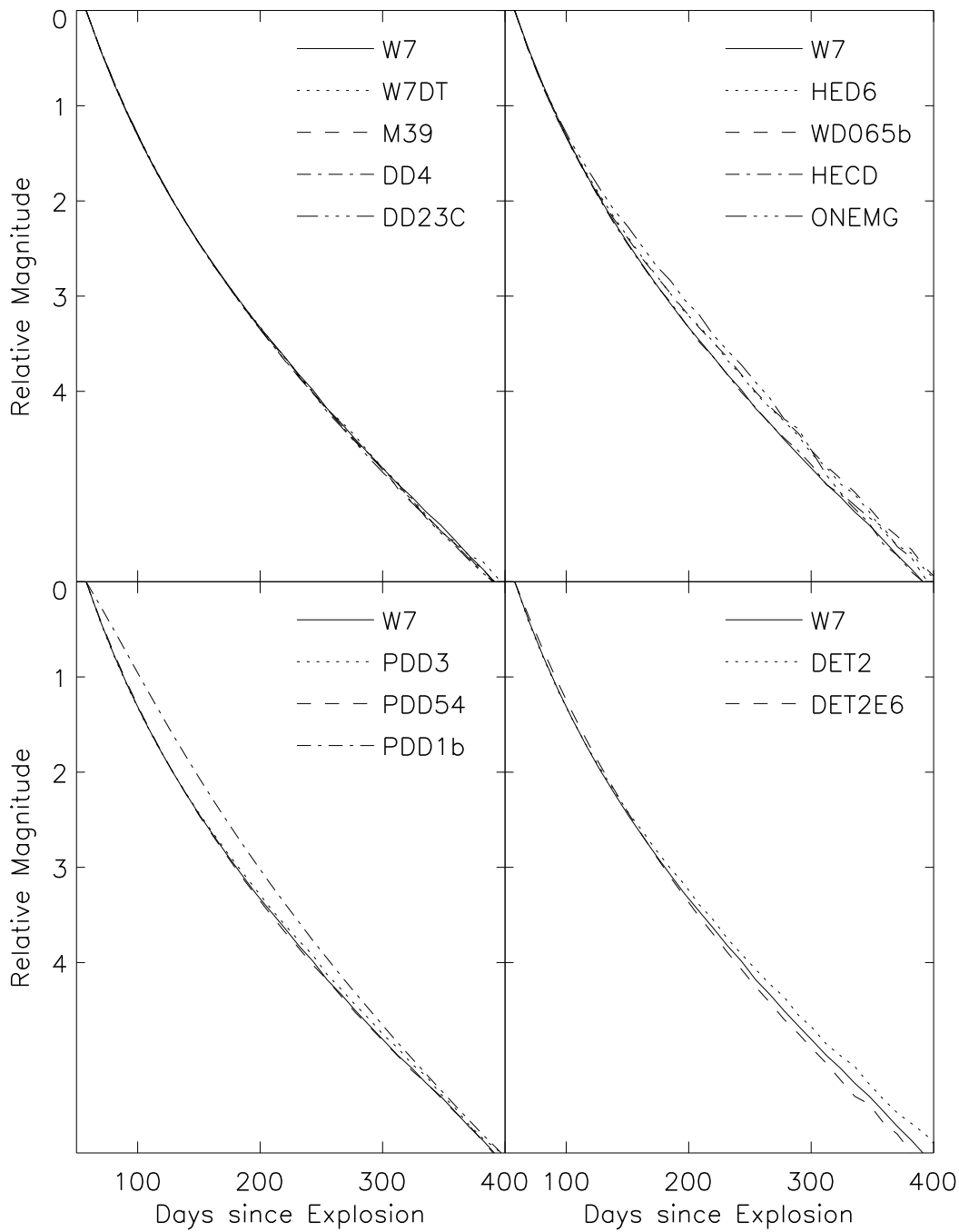}
\vspace{-0.5cm}
\caption{The family of bolometric light curves for models 
with radial magnetic field geometries and 1\% ionization, representing
the extremes of nickel mass, total mass and kinetic energy. The
models are grouped according to explosion type. The curves
are normalized to 60$^{d}$ to show the variation of the
  shape of the curves. The low-mass models remain brighter than W7
due to the onset of the positron-dominated epoch. PDD1b remains
brighter than W7 due to the efficient trapping of gamma photons out
to 300 days. 
 \label{fam400} }
\end{figure}

In figure \ref{gamma} we show the luminosity due only to the gamma
deposition. As expected, the total luminosity roughly traces the
amount of $^{56}$Ni produced. The steepness of the early
decline measures the mass overlying the Ni. The models then flatten,
with slopes becoming nearly equal. In figure \ref{fam400} we add 
the positron contribution to create bolometric light curves for
radial field geometries, assuming 1\% ionization. The
curves start at and are normalized to 60$^{d}$ to show the evolution of
their shape from 60$^{d}$ -400$^{d}$. There are often many observations
during this epoch, so the different shapes provide a test of whether the
model (regardless of field characteristics) fits the light curve. 
One interesting feature is the steepness of the
Chandrasekhar mass models, in which the $^{56}$Ni is covered by
a large overlying mass. 
We interpret this steepness to be due to the delayed onset of the 
positron-dominated epoch. The low mass models enter this epoch earlier, so
they flatten toward the decay line. Table \ref{table1} shows
the day when gamma deposition falls to equal the positron deposition.
HED6, WD065, ONeMg, DET2 have relatively little mass and transition before
180$^{d}$; PDD1b and DET2ENV6 have much more overlying mass and
transition at 300$^{d}$ or later. PDD1b thins to gamma photons so slowly,
that it remains as bright as the low mass models due primarily to gamma
deposition. Other than the extreme model PDD1b, shallow 
declines after 60$^{d}$ suggest
models with less mass overlying the $^{56}$Ni, steep declines suggest more  
overlying mass. This trend is also evident in the light curves  
calculated by 
RLS, who parametrize the slope at various epochs in their
Table 3.  The separation of WD065 and ONeMg from HED6 after 200$^{d}$ is due
to the earlier survival of positrons in WD065 and ONeMg.

\begin{figure}
\plotone{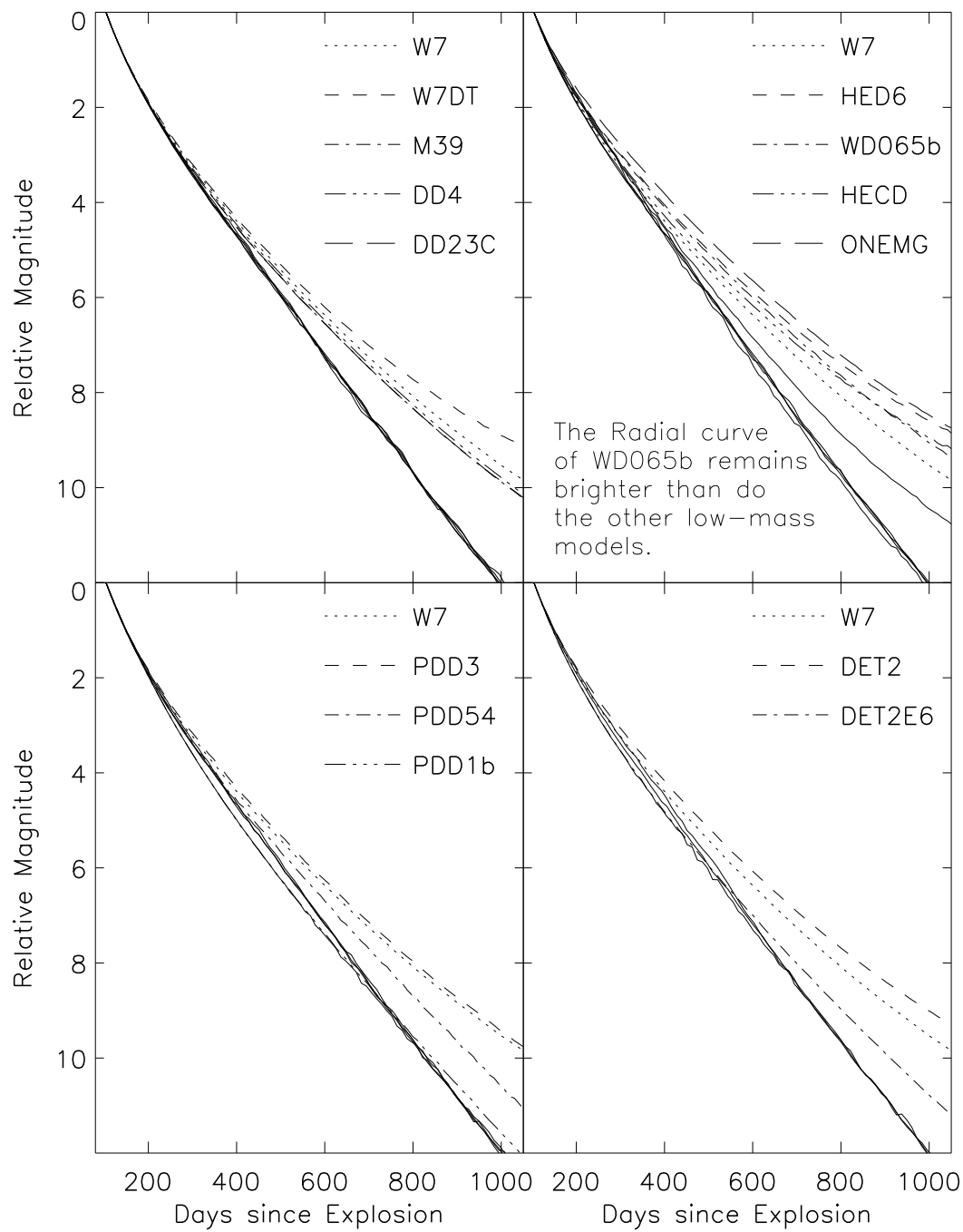}
\vspace{-0.5cm}
\caption{
Similar to figure \protect\ref{fam400} with the curves extended to
1000$^{d}$, and both radial (solid lines) and trapping 
field geometries (various other lines) are shown. 
 The radial curves form an approximately homogeneous
 group, the trapping curves flatten
according to the $^{56}$Ni location and the 
velocity structure of the nickel-rich zones. 
\label{fam1000} }
\end{figure}

Figure 
\ref{fam1000} is an extension of figure \ref{fam400} to 1000$^{d}$,
but including the curves from positron trapping scenarios.
This epoch emphasizes the differences in positron transport between the
field geometries. The most noticeable feature is that the radial models are
approximately equivalent and steep,
 whereas the trapping models flatten according 
to the percentage of $^{56}$Ni  produced in the outer portions of the SN.
RLS mention that the late light curves of trapping field
scenarios flatten relative to instantaneous deposition,
 but state that the effect is small. Our
results show 
the effect to be large. Thus, 
the 400$^{d}$ -800$^{d}$, positron-dominated epoch is a diagnostic
of field characteristics, not of model types.
Also apparent in this figure is that the ``massive'' models 
 show separation between the field
scenarios much later and to a lesser degree than the rest of the models.
Massive in this sense refers to a large mass of slow ejecta overlying the
$^{56}$Ni-rich zones.

The variety of shapes at early epochs and then the dramatic separation
between the predictions of positron trapping in a turbulent field 
geometry and positron escape in a radial or weak field show that
60$^{d}$ and later bolometric light curves yield a wealth
of information as to the structure and dynamics of the SN ejecta.

\section{Comparison With Observed Light Curves}

Ideally, to probe the SN structure using light curves, 
 model-generated bolometric light curves are compared with 
observed bolometric light  curves. Observed bolometric light curve
reconstructions, to date,
are at best based on measurements in the U,B,V,R,I bands, with some information
from the J,H and K bands. As the SN dims, the photometric uncertainties
increase and for many light curves, the number of bands 
observed decreases, leading to less accurate 
 bolometric reconstructions.
We use
the available bolometric light curves: SN 1992A 
to day 420 (Suntzeff 1996),
 SN 1991bg to day 220 (Turratto et al. 1996), 
 SN 1972E 
to day 420 (Axelrod 1980)
\footnote{We do not use the 720 day data point because we
consider the extrapolation of an entire bolometric light curve from a
1200$\AA$ wide spectrum to be too uncertain}, SN 1989B to
day 
135, and SN 1991T to day 108 (Suntzeff 1996).
 The best epoch to observe positron transport effects is during 
the 500$^{d}$
-900$^{d}$ range; only SN 92A and SN 72E extend close to that epoch.
Nonetheless, the SN 91bg, SN 89B and SN 91T light curves 
are valuable in checking the fit of a specific model during the
gamma-dominated phase.
For each model we consider for a particular event, we simply fit
the calculated light curve to the ``observed'' bolometric light
curve. Thus we avoid the uncertainties in 
distance, extinction, and bolometric correction.
We then show the later, positron-dominated light curves relative
to this fit for comparison to the more limited data in one or a few
bands.
 RLS fit models to the 
bolometric luminosities of SN 1991bg and
SN 1992A, Mazzali et al. (1996)
 fit SN 1991bg. In sections 5.1 5.2 and 5.4 we will discuss our
results in relation to theirs.

When there is insufficient data to reconstruct bolometric light curves we
compare model bolometric shapes to individual band photometry. This 
assumes that at late epochs there is little 
color evolution. We must be able to rule out any increasing shift of
the luminosity into unobserved bands.
Lira (1998) showed that the collective tendency is for the 
 color indices stop evolving after
day 100-120$^d$.
In figure 10 we show
the U, B, V, R, and I band variations of six of the SNe used in this study.
B-V peaks around t$\simeq$30-40$^d$, then decreases to approximately zero
near day 100 (except SN 1991bg).
At the (B-V) peak, the V band contains most of the energy and then declines
to B-V $\simeq$ 0 with the B band a potentially important contributor.
The V band is the 
best single band to trace the supernova bolometric light curve.
The five bolometric light curves permit us to estimate the error 
introduced by fitting with only the V band.
As shown in figure 11, the inaccuracies in using the V or B band data
for the bolometric luminosity are $\leq$0.2$^m$ during 
the 60$^d$-120$^d$ epoch. 
The similar procedure in comparing model-generated bolometric 
light curves to band photometry was also employed
by CAPP who used
V band data for every epoch.

Theoretical arguments about color
evolution have been made by Axelrod  and Colgate (1996),
who disagree as to whether collisional or radiative
processes dominate the emission from the recombination cascade. 
The competition between collisional and radiative processes hinges
upon
three factors: level of ionization (temperature in the collisional
scenario), density, and atomic cross sections.
Fransson \& Houck (1996) addressed these 
issues, calculating multi-band model
light curves. The 
technique worked well for the type II SN 1987A, but the 
type Ia light curves 
generated from the model DD4 showed a sharp decrease in U,B,V,R,I
around day 500 due to the infrared catastrophe (IRC),
and were inconsistent with SN 1972E. Why the IRC does
not apparently occur in SN Ia is something of a puzzle.
 The 
model
DD3  contains more $^{56}$Ni (0.93 M$_{\odot}$ versus 0.62
M$_{\odot}$ for DD4) and thus maintains a higher level of ionization.
This delayed the onset of the IRC, but  
still could not fit the observed light curves of SN 
72E. It is important for our discussion that any onset of the IRC
is abrupt; this does not lead to a gentle decline of the light curve
as we find for positron escape.
Fransson \& Houck (1996) also considered clumping in the ejecta.
It might delay the onset of the IRC for the inter-clump regions,
but it hastens its onset in the clumps. It remains to be seen if
spectra modeled with clumping will reproduce the observed 
spectra.

The atomic cross sections provide a third explanation for the lack of
color evolution. If the radiative transition probabilities are underestimated, 
 the radiative scenario might dominate. Fransson \& Houck increased
the recombination rates by a factor of 3 to model the spectra; perhaps
additional adjustments are required.
Few of the 
 SNe observed at late epochs show convincing color
evolution. Two of the three SNe that continue to evolve after 
120$^{d}$, 1986G and 1994D, seem to settle into  constant color indices later.
All the evidence suggests to us that the V (and probably B) band tracks the
bolometric luminosity of SN Ia at late times.
We emphasize the SNe for which there are at least two measured bands  for
most of the observed light curve.

\section{Comparison of Models to Individual SNe}

The amount, type, and quality of data varies among SNe. 
A summary of the observations used in this study are listed in 
table \ref{sumobs}.
A wider range of model fits to SNe is shown in Milne (1998).
We primarily consider  SN Ia models shown by others to describe well
the early light curves and/or spectra of particular
well-observed events, calculating their light curves
to late times.  
Models are considered for a given event if they can reproduce some combination of
the following features: distance-dependent peak luminosity estimates,
rise-time to peak luminosity and peak width, early spectral features and
light curve shapes for multiple color bands.
Generally the early data available do not uniquely define the model
parameters, or even the basic type of model. 
For some cases, the suggested models are
quite different. An example of this is SN 1991bg, which is fit with two
low-mass models (an 0.6 M$_{\odot}$ AIC and a 0.65 M$_{\odot}$ HeDET),  
a 1.4 M$_{\odot}$
pulsed model and a 1.4 M$_{\odot}$ deflagration model.
Our first test is whether
a model can fit the earlier nebular light curve, which probes the 
transition from gamma  to positron domination.  We then 
 examine whether models that pass this first test can
determine the magnetic field configuration  and degree of ionization.
Thus, this study is able to add another
constraint to the SN fitting puzzle, as well as determining if positrons 
escape from SN ejecta. 

\subsection{SN 1992A in NGC 1380}

SN 1992A occurred in the S0 galaxy NGC 1380 and was observed extensively.
 It is often held up as an 
example of a normal SN Ia. Extinction appears to be minimal, making SN 1992A 
an excellent candidate for photometric analysis.
 SN 1992A is one of three SNe treated
in both RLS and CAPP. We show the fits of three 
models to SN 1992A: DD23C, M39 and HED8. 
DD23C was fit on the recommendation of Peter H\H{o}flich (1998),
and also  because Kirshner et al. (1992) fit the spectral data with 
a modified version
of DD4, which is similar to DD23C. 
RLS combined the distance estimates to NGC 1380 with Suntzeff's UVOIR 
bolometric estimation to suggest that the peak bolometric luminosity 
was between 42.65 and 43.00 dex. 
 This suggests that a sub-luminous model might be  
required.\footnote{A recent study by Suntzeff et al. 1998 suggests that 
NGC 1365 (treated to represent the center of the Fornax cluster) may 
be a foreground galaxy in the cluster. If this is the case, then SN 1992A 
may be only mildly sub-luminous, and at a distance in agreement with 
Branch et al. 1997.}
 M39 is a delayed detonation that has a peak bolometric luminosity
 of 43.06. We choose it as a compromise between the delayed detonation
 scenario suggested by the spectra, and the low luminosity suggested by distance
 estimates.\footnote{We note that the B-V color of M39 is too red to fit
92A according to the model generated light curves of H\H{o}flich, even
with zero extinction.}
CAPP listed the distance to NGC 1380 as 16.9 Mpc, with
 E(B-V) = 0.00$^{m}$, and fit the data with a 1.0 M$_{\odot}$ model which 
 produced 0.4 M$_{\odot}$ of $^{56}$Ni. We 
 instead use the similar model HED8. 

 Figure \ref{bol92a} shows the fits of DD23C and HED8 to the
bolometric light curves of SN 1992A. For this object the bolometric
light curves are reconstructed to late enough times that we can
use them directly to study positron energy input.
 Assuming a 0.025 dex 
 uncertainty for each data point, 
DD23C provided the best fit, varying only the overall amplitude, 
of the suggested models 
 in the 55--420$^{d}$ epoch,
 fitting with 81\% confidence for the radial 
 curve. The trapping version of that model was rejected at the 99.91\%
 confidence even when renormalized. The numerical results are shown in 
table \ref{table2} for 1\% ionization, triple ionization with a 
radial field geometry, and for 1\% ionization with a trapping field. 
 The low mass model, HED8 did not fit above
the 10$^{-4}$ level for any scenario. 
 RLS also fit a 0.96 M$_{\odot}$ model to SN 1992A. Our results are 
 consistent with theirs in that the models are brighter than the data after 100 days. 
  The CAPP fit for 7 $\leq$ $\kappa_{e+}$ $\leq$ 10 remained too bright from 
  20$^{d}$--320$^{d}$, also in agreement with our results. The best
fitting class of models were the delayed detonation models; 
  W7, DET2ENV6
 and PDD54 were the only models other than delayed detonation models
 to fit at better than the 20\% 
  confidence level.  

We show the fits of two models 
to the V data, DD23C and M39, in figure 
\ref{v92a}. 
 Both models follow the falling luminosity better with
positron escape than with trapping. Fitting time-invariant
ionization scenarios to the V data with the published uncertainties, 
none of the models provided
a fit better than $\chi^{2}$/DOF=18. 
The numerical fits to the V band data (table 2)
show that with the scatter in excess of the stated uncertainties, none 
of the models provide statistically acceptable fits. 
 In absolute terms, our goodness-of-fit
statistics are questionable, but the fits 
are better for radial scenarios for almost every 
model.\footnote{The model DET2ENV6 fits the late V band data well (but less so 
the bolometric data) and may warrant further investigation.}

  The other delayed detonation models 
yielded fits similar to DD23C and M39 for the 
  V data, demonstrating that the observation of positron escape at late 
  times is not strongly model-dependent. The 
  928$^{d}$ point might herald the onset of another source
  of power.
Possibilities include other radioactivity, such as $^{22}$Na or $^{44}$Ti
(but not $^{57}$Co, which is too weak), recombination energy lagging  
earlier, higher  ionization input (Clayton et~al. 1992, Fransson \& Kozma
1993), and positrons encountering circumstellar material, as well as
others. In summary for SN 1992A:
  delayed detonations are 
  the most promising model types, all of which overproduce the late light
V light curves without escape of positrons.

  \subsection{SN 1991bg in NGC 4374}

   SN 1991bg occurred in the galaxy NGC 4374 and is the best 
   observed member of a class of 
   sub-luminous SNe that 
   have a fast decline from peak luminosity. SN 1991bg appeared to be 
   very red at peak suggesting either significant extinction ($\sim$  
   0.7$^{m}$), an intrinsically red SN, or both. The fact 
   that SN 1991bg and SN 1986G continued to show color evolution 
   after 120$^{d}$ 
   suggests that they were intrinsically different than normal SN Ia.
   We fit SN 1991bg with three models that have been suggested to explain 
   this SN; WD065, PDD54 and ONeMg, as well as a fourth model, W7. 
   The models WD065 (a helium detonation) and ONeMg (an accretion 
   induced collapse) decline faster from peak luminosity relative to 
   1.4 M$_{\odot}$ models because they have less 
   overlying mass (but similar velocity structures) leading to a 
   lower optical thickness. Ruiz-Lapuente (1993) fit WD065 to the maximum and
nebular spectra, and later (RLS) to the bolometric light curve.
PDD5 is a pulsed delayed-detonation model that 
   produces very little $^{56}$Ni, and maintains a lower level of
   ionization (relative to W7 and normally luminous models) which 
   decreases the free-free opacity and thus the overall opacity. 
   H\H{o}flich (1996)
 fit PDD5 to B,V,R,and I band data out to 75$^{d}$, suggesting
the distance to NGC 4374 to be 18 $\pm$ 5 Mpc with E(B-V)=0.25$^{m}$.
 PDD5 is unavailable, so we use the similar PDD54. 
   W7 was included because it provides the best fit of all 
   models, it has not been suggested by other authors.
   The fits for the four models to the bolometric data 
 are shown in figure \ref{bol91bg}; fits to the V data is shown in 
figures \ref{v91bg1} and \ref{v91bg2}. 
    Present in the B and V data is a disagreement 
   after 140$^{d}$ between the data taken by Leibundgut et al. (1993)
 and that taken by Turatto (1996),
 the Turatto data suggesting the SN was 
   fainter.\footnote{The third data set of Filippenko tends to
  confirm the measurements of Turatto, 
  but do not extend beyond 140$^{d}$.} 
   The bolometric light curve was calculated by Turatto et al. (1996) from the 
   photometric data, and strongly relied upon the B and V bands. 
   No models were able to reproduce the steep decline of the Turatto 
   data. Placing 0.04 dex error bars on the bolometric data after
  50$^{d}$ and ignoring
   the data after 140$^{d}$, PDD54 provided the best overall fit
 at the
27\% level. WD065b fit below the 2\% level, ONeMg
below 10$^{-5}$. 
Every model remains too bright to fit the bolometric data out to 250$^{d}$,
with no model fitting
above the 1\% level. 
   
   The B and V band data shows the models WD065 and ONeMg to be bright 
   from 140$^{d}$--250$^{d}$,
   too bright to fit either set of data. PDD54 and W7 
   were able to fit the Leibundgut data within the uncertainties. RLS fit 
   WD065 to the bolometric data, invoking the weak-field scenario. For 
our calculated WD065 light curve, 
 the weak-field scenario was fainter than the radial scenario, 
   but the light curve remained too bright to fit the data. The dashed line
   shows the prediction for 100\% transparency to positrons (zero 
   deposition of positron kinetic energy). This line fits the data, but 
   there is no physical justification for this extreme scenario. Our results 
   agree with the results of CAPP, who tried to fit a 0.7M$_{\odot}$ model
   that produced 0.1M$_{\odot}$ of $^{56}$Ni and determined that only 
   zero positron deposition ($\kappa_{e+}$=0 in their terminology) would 
   approximate the data. For ONeMg, the slope of our bolometric light 
   curve agrees with Nomoto et al. (1996) from 60$^{d}$--90$^{d}$. It is the 
   120$^{d}$--450$^{d}$ epoch that excludes ONeMg as an acceptable model.
   Mazzali et al.(1996) tried to reproduce the spectra of SN 1991bg with a 0.62
M$_{\odot}$ version of W7, and as a
by-product generated a bolometric light curve out to 220$^{d}$
post-explosion. Their light curve assumed positron trapping and was 
too bright to fit the data after 110$^{d}$. They argued that  this epoch
is too early to expect positron escape and suggested that there is
unseen emission leading to an erroneously low bolometric light curve.  
Our light curves for the low-mass models WD065b and ONeMg have the same
characteristics as theirs, and positron escape is insufficient to
explain the low luminosity in the 110$^{d}$--220$^{d}$ epoch. 

The inability to fit any model to the Turatto data tempted us to favor the 
trend emerging from the Leibundgut data over the Turatto data. But, the 
light curve from SN 1992K forbids that action. SN 1992K has near peak 
spectral and photometric properties similar to SN 1991bg, and the B and V 
data follows a SN 1991bg-like shape. The V data extends out to $\sim$155$^{d}$.
The existence of a second example of this steep decline forces us to 
conclude that there exists a sub-class of type Ia SNe for which we are 
unable to fit the light curves with the models in our possession without 
invoking a larger extinction than suggested. The 
flatness of low mass models during the positron-dominated phase
 suggests that that class of models possess the 
opposite tendency from that required. The solution to this problem 
remains unclear. 
   
   \subsection{SN 1990N in NGC 4639}

   SN 1990N was unusual in that Co lines were detected 
   earlier than existing models predicted and the intermediate mass 
   elements had higher velocities (Leibundgut 1991).
 The near-maximum and post-maximum 
   spectra were normal, suggesting normal models such as W7 could 
   explain them. The late detonation model, W7DN, was created to 
   fit the early spectra and transition to W7. 
   This was 
   accomplished by having a deflagration accelerate into a detonation 
   at M$_{r}$=1.20M$_{\odot}$. Yamaoka (1992) discussed how the 
   extra $^{56}$Ni modifies the rise to peak light to fit SN 1990N.
   H\H{o}flich (1996) 
   fit
   DET2ENV2/4 and PDD3 to the multi-band photometry of SN 1990N, for
   a distance to NGC 4639 of 20 $\pm$ 5 Mpc. The 
   multi-band coverage is very good, but a bolometric light curve has 
   not been published. There was an unfortunate gap in the light curve 
   from 70$^{d}$--200$^{d}$, as the SN was too close to the Sun for 
   observations, hampering our ability to differentiate among
   model types.

   Figure \ref{bv90n} shows the models W7DN and PDD3 fit to the B and V 
   data.
 Both models fit the 190$^{d}$--300$^{d}$ data well and then
 show a steepening consistent with positron escape. 
    The normally 
   luminous DD and PDD models provided similar fits, with all models
   giving worse fits for positron trapping.

\subsection{SN 1972E in NGC 5253}

SN 1972E was not observed pre-maximum and thus the models are not 
well constrained. We assume the explosion date to be
JD2441420 (Axelrod 1980). The photometry out to $^{+}$169$^{d}$ 
is photoelectric, with
 later photometry derived from spectra (Kirshner \& Oke 1975). 
The bolometric light curve published by Axelrod was also generated by a 
model fit to the Kirshner spectra. We fit three models to 72E: W7, M35
and HED8. RLS fit W7 to the bolometric 
data, invoking a transition from turbulent confinement (with
B$_{o}$=10$^{5}$Ga) to weak-field 
trajectories around 500$^{d}$ 
Their W7 light curve fit well before 500$^{d}$. The transition was 
discussed conceptually by RLS, but not modelled. 
 H\H{o}flich (1996) fit the model M35 to the
multi-band photometry of 72E, suggesting the distance NGC 5253 to 
be 4.0 $\pm$ 0.6 Mpc. The model HED8  is shown because
of 
the agreement with the bolometric light curve. All three 
models are consistent with the $^{56}$Ni mass suggested by the nebular 
spectra, 0.5--0.6M$_{\odot}$ (Ruiz-Lapuente \& Lucy 1992). Colgate (1996)
suggests a 0.4 M$_{\odot}$ model to explain 72E, this possibility was 
not investigated due to the lack of a suitable model. 

Figure \ref{72e} shows the fits of 
HED8, W7, and M35 to the B, V and bolometric data.
None of these models can be rejected, but all fit
better with the radial field scenario and significant
positron escape. None of the models 
fit the 700+$^{d}$ data point, with all remaining too 
bright. 

\subsection{SN 1991T in NGC 4527}
 
SN 1991T was unusual in the width of the luminosity peak and in the 
absence of SiII and CaII absorption lines (Filippenko et al. 1997). 
It is the prototype of superluminous type Ia SNe. It was 
also an example of a SN whose late emission was overwhelmed
by a light echo, as shown by Schmidt et al. (1994). Suntzeff 
(1996) 
derived a bolometric light curve to 87$^{d}$ post-B maximum 
against which we tested models. There was also
a gap in observations as the sun was too near the SN. 
We show the fits of two models to SN 1991T, the late-detonation W7DT, 
and the superluminous helium 
detonation HECD.
The  
model W7DT was proposed by Yamaoka et al. (1992) 
to explain the lack of intermediate 
mass elements in the pre-maximum spectra, followed by a normal spectrum 
post-maximum. Nomoto et al. (1996) modelled the bolometric light curve for 
W7DT and scaled it to fit the V data. 
 H\H{o}flich (1996) suggested PDD3 and DET2ENV2 based upon fits  
to multi-band photometry, suggesting the distance to NGC 4527 to be 
12 $\pm$ 2 Mpc. Liu et al. (1997) argued that the 
superluminous 1.1 M$_{\odot}$ helium detonation, SC1.1 (which produces 
0.8 M$_{\odot}$ of $^{56}$Ni), fit the nebular spectrum (301$^{d}$) 
better than did the models W7, DD4 and SC0.9. To fit SN 1991T, Liu et al. 
(1997)  assumed a 
distance of 12--14 Mpc and E(B-V)=0.0$^{m}$--0.1$^{m}$. We 
approximate 
SC1.1 with HECD, a 1.06 M$_{\odot}$ helium detonation that produces 
0.72 M$_{\odot}$ of $^{56}$Ni. 
Pinto \& Eastmann (1996) fit DD4 (with pop. II
elements added) to the B and V data for SN 1991T out to 60$^{d}$, suggesting
a distance of 14 Mpc for E(B-V)=0.1$^{m}$.

RLF estimated 
the $^{56}$Ni production to be 0.7--0.8 M$_{\odot}$ 
of $^{56}$Ni using nebular 
spectra; 
Bowers' (1997) $^{56}$Ni production (when adjusted as 
explained in section 4) is 0.7$\pm$0.2 M$_{\odot}$. W7DT and HECD are  
both consistent with this, while PDD3 slightly 
underproduces $^{56}$Ni. 

Figure \ref{91t} shows the fits  of W7DT and HECD to the V data.
 The late V data was fit by adding a constant
light echo component. Both models show that the radial field
scenario with positron escape provides better fits. 
The dashed lines show the radial
light curves without an echo, the dot-dashed lines show the trapping 
light curves without an echo. The trapping curves remain too bright
during the 200$^{d}$--500$^{d}$ epoch 
even with no contribution from a light echo, arguing strongly against
trapping. The 1.1 M$_{\odot}$ and 1.4 M$_{\odot}$
 models have similar late light curves, 
suggesting that they can not resolve the ambiguity between 
Chandrasekhar and sub-Chandrasekhar mass models for superluminous SNe.
 The fact that the radial curve explains the data with no
echo contribution until 470$^{d}$ perhaps suggests that the
light echo that ``turned on'' after 450$^{d}$, an effect that could
be explained by the SN light sweeping through a dust cloud. 
The asymmetry of the image of the SN 1991T light echo would be consistent 
with reflection off of discrete cloud(s) (Boffi et al. 1998). 
 Unfortunately, the light echo interrupted the positron dominated 
epoch before the optimal time to observe positron escape ($\sim$600$^{d}$).
Whereas it may be possible to eventually account for and subtract out 
the contributions from a light echo, we do not fit 
light curves into the light echo phase other than to demonstrate the 
phenomenon.

\subsection{SN 1993L in IC 5270}

SN 1993L was observed by CAPP and fit by the same model used 
for SN 1992A. The strength of the SN 1993L data is the
continuous multi-band photometry extending to beyond
500$^{d}$. The data were published without uncertainties. There are no 
published spectra, so the best model was selected entirely by
the fit to the B and V data. An additional complication is that the SN 
was discovered after maximum, which eliminated that discriminant.
CAPP assumed the explosion date to be JD 2449098, 
the distance to be 20.1 Mpc and 
 the extinction to be 
E(B-V)=0.75$^{m}$.   
With these assumptions, SN 1993L appears to be  quite similar 
to SN 1992A. CAPP noted two differences between the two SNe; SN 1993L had a 
slower nebular velocity and a higher degree of V band curvature from 
100$^{d}$--500$^{d}$. We note the additional difference that the SN 1993L 
light curve was dimmer from 25$^{d}$--55$^{d}$ and again from 100$^{d}$ 
to the cross-over at 400$^{d}$. Explaining these features 
is difficult, but we note that some of
the PDD and DD models studied by 
H\H{o}flich \& Khokhlov show these features. 

Assuming the photometry errors to range from 0.1$^{m}$---0.5$^{m}$ from 
60$^{d}$--550$^{d}$, only  PDD1b was excluded
at the 99\% confidence level, all other models and scenarios were 
consistent at the 30\% level or better. 
 We show in figure \ref{bv93l} the fits of HED8 and DD4 
 to the B and V data.
The radial curves for 
both models yielded similar fits within the scatter.
For HED8, the trapping curve remains much too bright to fit the data..
SN 1993L does not provide strong evidence of positron escape, but the radial
light curves fit at least as well as trapping curves for any models tested.

\subsection{SN 1937C in IC 4182}

SN 1937C reached a peak B  magnitude of 8.71$^{m}$ and was detected on 
photographic plates for over 600$^{d}$. Branch, Romanishin \& Baron
 (1996) have argued that  
that its spectral features are those of normal SNe Ia, while Pierce \& 
Jacoby (1995) argued
that SN 1937C is similar to SN 1991T and thus superluminous. 
H\H{o}flich (HK96) suggests N32, W7 and DET2 as acceptable models,
and a distance of 4.5 $\pm$ 1 Mpc. 
 We use the data of Schaefer (1994),  which is a
re-analysis of photographic plates from many observers, principally
Baade (1938) and 
Baade \& Zwicky (1938). Pierce and Jacoby (1995) also
re-analyzed photographic plates of SN 1937C obtaining a higher value for 
B$_{max}$=8.94$^{m}$.

 Figure \ref{b37c} shows the fits  of  
DET2 and W7 
to the B data of SN 1937C.  
The 1\% ionization, radial-field 
light curves for DET2 and W7  fit the data well 
at all times, giving the
best fits ($\chi^{2}$/dof=1.6) of any model (and any field scenario). 
The field free scenarios for DET2 and W7 are 
nearly identical to the radial
field scenarios. 
The light curve for SN 1937C extends late enough with high-quality data
to convincingly trace a shape after 400$^d$ consistent with positron escape.
We also suggest that models with masses lower than 1.4 M$_{\odot}$ and
normally-luminous pulsed-delayed detonation models appear
to fit better than do delayed detonation models. 
We note that the same conclusions are reached if the original Baade and
Zwicky data is used, or the re-analyzed data of Pierce and Jacoby 
is used (Milne 1998).

\subsection{SN 1989B in NGC 3627}

     SN 1989B  occurred in NGC 3627 and had considerable extinction. 
   An additional complication is its location in the spiral 
   arm of the host galaxy, giving considerable background 
   contamination. The bolometric light curve of Suntzeff 
(1996) 
   shows SN 1989B to be similar to SN 1992A, 
  but remaining brighter than 
   SN 1992A from 90 days onward. HK96 suggested the models M37 
   and M36 based upon fits to the multi-band photometry; we fit M36 to
  B,V and bolometric data for SN 1989B. The distance suggested by HK96 
is 8.7$\pm$3 Mpc.
   
   Figure \ref{89b} shows the fits of M36 and ONeMg light curves to
the bolometric and V data from SN 1989B. 
   The late bolometric 
   and the later B and V band light curves 
   remained too bright to be fit by M36. HK96 fit the model M37
 to this same data, their V light curve remaining bright enough
to fit the data. Our light curve for M36 is 0.5$^{m}$ dimmer than the 
HK96 estimate for M37 at 365$^{d}$. H\H{o}flich \& Khokhlov cautioned against
over-interpreting the latter portion of their light curve, but as none
of the delayed detonations tested in this study approached the
brightness of their M37 light curve, further investigations may
be warranted.  The only model able to
  reproduce the light curve beyond 300$^{d}$ is the 0.6 M$_{\odot}$
model, ONeMg. To date, there has
been no suggestion that SN 1989B was substantially 
subluminous; in fact, it has been
considered a relatively normal SN Ia. 
The possibility  that SN 1989B was produced by a
low-mass WD can be tested, especially when the distance estimate is improved.
If this
explanation is correct, this SN is the singular example of a light curve
best fit with positron trapping. Another
explanation is that 
   the late light curve was affected by a light echo or another effect
 of the complicated background subtraction. There are two  reasons 
   to believe that a light echo may have been present. The 330$^{d}$
   spectrum  showed more continuum emission at 6000$\AA$ than 
   is typically  seen in the nebular spectra of SNe Ia.     
   In addition, 
   there was strong Na-D absorption from the
   host galaxy, an indicator of foreground dust. If the dust were 
   near enough to, or surrounding the SN, a light echo may have  
   been produced. As good spectra exist,
   this question is also potentially solvable. 
   The existence of two dramatically different explanations for the late
light curve of SN 1989B means the subtle effect of 
    positron escape can not be clearly seen or ruled out.

   \subsection{SN 1986G in NGC 5128}

  SN 1986G was well observed, but occurred in the dust lane of NGC 5128 
  and suffered from considerable extinction. SN 1986G had a narrow peak and 
  slow $\lambda$6355$\AA$ Si lines, suggesting that it was 
intermediate between normal SNe Ia and SN 1991bg. 
 The B-V color index continued to evolve 
  120$^{d}$ after the explosion, a feature also seen in SN 1991bg, but not 
  observed in normal SN Ia.\footnote{It is important to note that the 
B-V index appears to approach zero by the last observation, a reversal 
of the 100$^{d}$--320$^{d}$ behavior.} 
  We fit the models W7, M39 and HED6 to SN 1986G. 
  HK96 fit W7  to  multi-band light curves to 
  80$^{d}$ post-explosion, suggesting the distance to NGC 5128 to be 
4.2$\pm$1.2 Mpc. M39 was used because the 
  $^{56}$Ni mass of M39 is in closer agreement with the RLF estimate of 
  0.38$\pm$0.03M$_{\odot}$. HED6  represents moderately
sub-luminous, low-mass models. 

  Figure \ref{bv86g} shows W7, M39 and HED6  
with the SN 1986G B and V data. 
 The Cristiani et al. (1992) data is shown without uncertainties. 
The B band is corrected  for 1.1$^{m}$ of estimated
extinction. The model curves
are normalized  at 120$^{d}$, when the B and V bands cross over.
All three models roughly reproduce the shape of the V light curve from
60$^{d}$--100$^{d}$. The B data is brighter than the V data after 120$^{d}$
 and can be fit by both W7 and M39. HED6 remains brighter than 
both W7 and M39, and provides a poorer fit to the data. 
Only the single point at 
425$^{d}$ is late enough to test the positron transport conditions.
This observation is consistent with radial or no magnetic field and 
positron escape, but with a realistic uncertainty it might not rule
out positron containment.

\subsection{SN 1994D in NGC 4526}

SN 1994D occurred in NGC 4526 and was observed by 
three groups until June 1994 and by Cappellaro et al. thereafter.
 We fit M36 
and HED8 for this object. The wealth of multi-band photometry and spectra
at early epochs allowed H\H{o}flich (1996) 
to tightly constrain delayed detonation models and
to conclude that  M36 was the best, at distance 16.2 $\pm$ 2
Mpc.
 Liu  et al. (1997,1998) found that SC0.9 (a 
0.9M$_{\odot}$ model which produces 0.43M$_{\odot}$ of $^{56}$Ni) 
fits the 301$^{d}$ spectrum. We use the similar model HED8. 
CAPP fit the V band data of 
94D with W7. As seen in figure \ref{fam400}, 
W7 and M36 are virtually indistinguishable, so we use M36.
The B-V index continued to evolve after 120$^{d}$, necessitating the 
inclusion of both B and V bands.\footnote{As with
 SN 1986G,  B-V  approaches zero at late times.} 
Figure \ref{bv94d} shows the fits of M36 and HED8 to the  SN 1994D B and V 
data. The 150$^{d}$--300$^{d}$ gap precludes our 
discrimination among model types. The scatter after 
300$^{d}$ prevents examination of the positron escape.
It seems the light curves can not rule out either delayed deflagrations or
He-detonations.

\subsection{Summary of Observations}

Ten SNe were analyzed, including  super-
and subluminous events, reddened and unreddened SNe, old ones recorded on
photographic plates and recent SNe recorded on CCDs. Of the ten, five
show strong evidence of positron escape (92A, 37C, 91T, 72E, 90N), 
three others are also consistent with significant
positron escape but somewhat ambiguous, (93L,
86G, 94D), and for two others it was not clear which model actually 
described the early light curve and should be tested for its positron
transport later 
(91bg, 89B). Only SN 1989B suggests the possibility that a trapping field may provide
a better fit, and there are complications with that interpretation.
As a group, the supernova light curves fall more quickly 
than models, which fit well at early times, extrapolated to later times
with all the positron kinetic energy deposited in and radiated by the ejecta.
This is consistent with the escape of a substantial fraction of the positrons
emitted by $^{56}$Co after one year.

Regarding even earlier times, for sub-luminous SNe the Chandrasekhar 
mass models fit the light curves better than low-mass models. For 
normally luminous SNe, there are not enough observations from 
60$^{d}$--400$^{d}$ to choose between the model masses. The superluminous
SNe can apparently be fit equally well by  high-mass He-detonation models
and nickel-rich Chandrasekhar mass models.

\section{Type Ia SN Contributions to the Galactic 511 keV Emission}

Table 3 shows the positron survival fraction of 
type Ia SNe  at 2000$^{d}$ and the resulting
positron yields for all the models treated in this study. The range of values
reflects the extremes of ionization fractions. The
lower values correspond to triple ionization, the higher values are
for 1\% ionization. The yields vary between the two extremes  
by roughly a factor of three. The observations suggest that 1\%
ionization typically fits better, so we will quote 2/3 of that
yield as a conservative estimate for each  model.  
For radial fields, the
Chandrasehkar mass models have a lower survival fraction when compared
with equally luminous sub-Chandrasehkar mass models, but
the larger nickel mass partially compensates for that fact. As a result,
for all but the ``heavy " models, the  
yields are not strongly dependent upon model mass. 
The radial scenarios have greatly enhanced positron escape 
yields compared to the trapping
scenarios. 

%

Positron escape is best observed best in light curves when its relative
effect is large, between 400$^{d}$ 
\& 1000$^{d}$, but
the positron yield in absolute terms is determined earlier: 80\% of the
escaping positrons do so between 178$^{d}$ -546$^{d}$ for W7 (94$^{d}$
-492$^{d}$ for HED8).
It is conceivable, but probably not common, that the trapping field could
apply until, say, 500$^{d}$, when the field magnitude decreased to the
point that the Larmor radius reached the ejecta radius and the object
would cross over to the field-free regime. The
``release time" for a positron of a given energy
is proportional to the initial field strength, which might 
vary over many orders of
magnitude. For one object we might incorrectly infer the positron
escape for this reason, but not likely for many.

The positrons that escape retain a significant fraction of their emitted
energy, as shown by CL and  
 in figure \ref{emit}, which compares the emission spectrum (dashed
line) to the escape spectrum (solid line).
 With energies near 400 keV, the
positrons have a considerable lifetime in the ISM, giving diffuse 
galactic 511 keV emission. 
 To estimate the rate of positron injection into the ISM, we
assume a  2:2:1 ratio of
normal:subluminous:superluminous SNe. Considering 
DD23C, PDD3, and HED8, a reasonable 
yield for normal SNe is 8 x 10$^{52}$ positrons.
For subluminous SNe,  PDD54, DET2ENV6 suggest
a yield of 4 x 10$^{52}$ positrons.
From W7DT, DD4, and HECD
(suggested by 91T) we estimate a yield of 15 x 10$^{52}$ positrons.
Employing the above ratio then gives a mean yield of 
 8 x 10$^{52}$ positrons per SN Ia.

This gives a flux  
 (4$\pi$ D$^{2}$)$^{-1}$ $\cdot$ y $\cdot$ SNR $\cdot$ $f_{e+ \gamma}$,
where D is the distance, y is the positron
yield per SN event, and SNR is the SN Ia rate, and 
$f_{e+ \gamma}$ is the 511
keV photons emitted per positron. Taking D=8 kpc, y=8 x 10$^{52}$ e+
SN$^{-1}$,
SNR=0.0032 SN yr$^{-1}$\footnote{This value was obtained
 from Capellaro et al. (1997)
and assumes  H$_{o}$=65 km s$^{-1}$ Mpc$^{-1}$. Hamuy \& Pinto (1998)
 suggest a slightly larger value, 0.0042 SN yr$^{-1}$.},
 and $f_{Ps}$=0.58 photons e+$^{-1}$
 (which corresponds to a
positronium fraction of 0.95), the flux is 6.3 x 10$^{-4}$ photons
cm$^{-2}$ s$^{-1}$.
An estimate of the uncertainty in this flux can be
obtained by inserting D=7.7 -8.5 kpc, y=3 -10 e+ SN$^{-1}$, SNR=0.003
 -0.06 SN yr$^{-1}$, and $f_{\gamma e+^{-1}}$=0.5 -0.65 (corresponding
to a positronium fraction of 1.0 -0.9) into the
formula. The flux then ranges from (1.2 -8.6) x 10$^{-4}$ photons
cm$^{-2}$ s$^{-1}$.
 The flux is on the order of the Galactic bulge component of the
511 keV flux as measured by OSSE. Purcell (1997) estimated the bulge
flux to be 3.5 x 10$^{-4}$ photons
cm$^{-2}$ s$^{-1}$ and the galactic plane flux to be 
8.9 x 10$^{-4}$ photons cm$^{-2}$ s$^{-1}$.  
 SMM, with a 130$^{\circ}$ FOV, (Share et al. 1988) 
measured the total flux from the general
direction of the galactic center to be 
2.4 x 10$^{-3}$ photons
cm$^{-2}$ s$^{-1}$; the type Ia SN contribution could be one-fourth of
that flux. A more exact treatment of the level of agreement between the
511 keV flux as mapped by OSSE and the spatial distribution of Ia SNe
is the subject of a forthcoming paper (Milne 1999).

\section{Summary}

   We calculate the $\gamma$-ray and the positron kinetic energy deposition
to produce ``UBVOIR'' bolometric light curves for the time when
the photon diffusion time (t$\geq$60$^d$)
is short for various models of type Ia SNe.
In calculating positron kinetic energy deposition into the ejecta,
we calculate it for particular extreme environments, they are:
radial magnetic field, turbulence magnetic field, and field free
geometries in a 1\% and triple ionization medium throughout the
evolution. 
The deposited energy rate is assumed to instantaneously appear as
``UBVOIR'' bolometric luminosity and is compared with several observed
bolometric luminosity when they are available or with B and V bands
observed luminosity.
In this work, we analyzed ten late time type Ia supernova light curves by
comparing the light curves with the calculated light curves.
It can be shown clearly, that 
all of the light curves except of SN 1991bg (explain below) require
positron kinetic energy deposition in order to give a reasonable
agreement. Without the energy deposition, it is quite obvious that
the light curve is too dim to explain the observe light curve at
t$\geq$200$^d$ (Colgate, Petschek, Kreise 1980; Capellaro et al. 1997;
Ruiz-Lapuenta \& Spruit 1997).

 We show $\gamma$-ray energy deposition at t$\geq$60$^d$ can hardly
be used to distinguished light curve shape between various models, 
except to differentiate between extreme models such as
between low (PDD1b) and high explosion energy models or
between low sub-Chandrasekhar mass (i.e. ONeMg) and 
Chandrasekhar mass (i.e. W7) models.
A similar situation also ensues in the epoch when positron kinetic
energy deposition is dominant and for the case of magnetic field in the ejecta
being radially combed outward.
In the case of turbulence magnetic field, the usefulness of the late time 
light curve to differentiate models is moderately improved due to
the effectiveness of ejecta to slow down and to absorb the
kinetic energy of positrons.

  The light curves of SN Ia are dominantly powered by 
the kinetic energy deposition
from $\beta^{+}$ decay of positrons after the nebula becomes optically
thin to gamma photons. 
Before and at the onset of this ``positron phase'' the
positrons have short lifetimes regardless of the field geometry, but as
the nebula becomes more tenuous positrons can survive and possibly
escape for favorable geometries. 
The longer the positrons stay as energetic particles, the larger is the
deficit of energy deposition relative to the full instantaneous
energy deposition. This effect is also observed when there are more
positrons escape from the ejecta.
Comparing late time light curves of suggested models 
from early light curve and spectra for  particular observed supernovae,
the deficit of luminosity relative to
instantaneous deposition can be seen in the B and/or the V band light
curves of SNe 1992A, 1990N, 1937C, 1972E, 1991T.
The phenomena give the evidence that positron take its time in 
depositing its kinetic energy and possibly moves quite far from its 
origin and even escapes from the ejecta.

Further detail comparisons of
the calculated light curves of positron energy deposition in
radial and tangled magnetic field configurations of type Ia models with
the observed V and B band light curves
show that the radial field configuration
(or synonymously the weak field configuration)
produces a better agreement than does the
tangled magnetic field configuration, which produces light
curves far too flat.
The agreement that requires radial field configuration is exhibited by
the observed
light curves of SN 1992A, SN 1937C, SN 1990N, SN 1972E, and SN 1991T.
Models with tangled magnetic field configuration produce a flatter
and brighter light curves than the radial field configuration due 
to the more energy deposition.
There are a number of potential sources which can flatten the curves
further, but steepening the curves such as by IRC
is difficult without color evolution,
which to date has not been observed in SN Ia light curve. 
These facts strengthen the argument
that positron escape occurs in SN Ia ejecta.

Fitting observed light curves of SN 1993L, SN 1994D, and SN 1986G  
give ambiguous results that it is not clear which of the 
radial or the tangled magnetic configuration model can give 
acceptable fit. 
The ambiguity is mostly due to the fluctuation in the observed
B and V band light curves. 
The strongest indication for the tangled magnetic configuration seems
to be shown by the SN 1989B observed light curve, however
there is a possibility that there is a contribution from a light echo
at late times.

SN 1991bg is somewhat a special case that not all SN 1991bg 
observed light curves agree with each other completely. 
Two light curves (Turatto et al. 1996; Filippenko et al. 1992) 
agree with each other, but no model that we have can give a 
reasonable fit to the light curves 
unless we unphysically impose no positron kinetic energy
deposition in the ejecta. 
The light curve observed by Leibundgut et al. (1993) can
be fit reasonably well with model PDD54 and W7, 
but we cannot determine the magnetic field configuration in the ejecta 
because of no observation done by Leibundgut et al. (1993) beyond day 200. 

Based on the shape of SN 1937C, SN 1972E, and SN 1992A's light 
curves, which all 
follow the predictions of the radial field configuration model,
we conclude that in these supernovae the positrons escape with the most
efficient transparency as if the positrons moved in a straight line
from their origins to the surface and deposit a minimum of 
their kinetic energy as
they escape. These supernovae show that type Ia SNe can be
a dominant source of the diffuse Galactic 511 keV line fluxes.

To better quantify the 
suggestion that type Ia supernova may be the main source of
positrons that produce the observed Galactic  511 keV line  fluxes.
We present the number or fraction of positrons that escape from
Ia models. The values are in agreement with the
values calculated by CL, who demonstrated that 
the radial magnetic field scenario
can easily provide the needed positrons to explain the Galactic
annihilation line fluxes.

The Galactic 511 keV annihilation line flux distribution
has a definite bulge component (Purcell et al. 1997).
SN Ia positrons from $^{56}$Co decays may contribute the majority of the
bulge flux and a sizable fraction of the entire galactic emission.
 As the number of SN Ia observed in spiral galaxies increase,
the spatial distribution of bulge and disk SNe will be better known. As
OSSE maps the galactic center region with increasing exposure and
spectral information, the level to which SN Ia positrons escape from SNe
may soon be known.

\acknowledgments

This work is from a dissertation which was submitted to the 
graduate school of Clemson University by P.A.M. in partial 
fulfillment of the requirements for the PhD degree in physics.
P.A.M. would like to thank P. Ruiz-Lapuente for access to models and
for emphasizing the use of bolometric data when available. We thank P.
H\H{o}flich for access to many models, giving assistance in locating
data and for suggestions of models to
test against individual SNe. The isotopically complete models provided
by K. Nomoto and S. Kumagai were important for many additional
applications. P.A.M. thanks D. Jeffery for spectroscopic insight and
assistance locating data. We thank P. Pinto for providing us with model
DD4 and for useful discussion on ionization fraction and plasma energy loss.
This work was supported by NASA grant DPR S-10987C, via sub-contract to
Clemson University.

\begin{deluxetable}{lcccccccc}
\footnotesize
\tablecaption{SN Ia model parameters. \label{table1} }
\tablewidth{0pt}
\tablehead{
\colhead{Model} & \colhead{Mode of } & 
\colhead{M$_{\star}$} & \colhead{M$_{Ni}$ } &
\colhead{E$_{kin}$}   & 
\colhead{t$_{B}$} &
\colhead{\.{E}$_{\gamma}$(60$^{d}$)} & 
\colhead{\.{E}$_{\gamma}$(100$^{d}$)} &
\colhead{T(\.{E}$_{e+}$=\.{E}$_{\gamma}$)} \\
\colhead{Name} & \colhead{Explosion} & \colhead{[M$_{\odot}$]} & 
\colhead{[M$_{\odot}$]} & 
\colhead{[10$^{51}$ erg s$^{-1}$]}  & \colhead{[days]} \tablenotemark{a} &
\colhead{[10$^{42}$ erg s$^{-1}$]} & \colhead{[10$^{42}$ erg]}  &
\colhead{ [days] \tablenotemark{b}} 
} 
\startdata
W7 & deflagration & 1.37 & 0.58 & 1.24 &14.0 &1.54 & 0.45 & 260 \nl
DD4 & delayed det. & 1.38 & 0.61 & 1.24 & 19.0&1.81 &0.54 & 250 \nl
M36 & delayed det. & 1.40 & 0.59 & 1.54 & 13.3 &1.67&  0.48 & 220 \nl
M39 & delayed det. & 1.38 & 0.32 & 1.42 & 14.3 &1.06& 0.32 & 240 \nl
DD202C & delayed det. & 1.40 & 0.72 & 1.33 &--- &2.02&0.57 & 235 \nl
DD23C & delayed det. & 1.34 & 0.60 & 1.17 &18.0&1.83 &0.57 & 245 \nl
W7DN & late det. & 1.37 & 0.62 & 1.60 & --- &1.63 &0.49 &230 \nl
W7DT & late det. & 1.37 & 0.76 & 1.61 & --- &1.76 &0.53 &205 \nl
PDD3 & pul.del.det. & 1.39 & 0.59 & 1.39 & 16.0&1.44 &  0.40 & 190 \nl
PDD54 & pul.del.det. & 1.40 & 0.17 & 1.02 & 12.3 &0.66 & 0.19 & 245 \nl
PDD1b & pul.del.det. & 1.40 & 0.14 & 0.35 & 11.8 &1.05 &0.45 & 520 \nl
WD065 & He-det. & 0.65 & 0.04 & 0.74 &13.5 & 0.09&0.02 & 170 \nl
HED6 & He-det. & 0.77 & 0.26 & 0.74 & 14.4&0.44 & 0.12 & 160 \nl
HED8 & He-det. & 0.96 & 0.51 & 1.00 & 13.7 &0.97 & 0.26 & 270 \nl
HECD & He-det. & 1.07 & 0.72 & 1.35 &--- &1.59 &0.46 & 200 \nl
ONeMg & AIC & 0.59 & 0.16 & 0.96 &15.0 &0.18 &0.05 &  130 \nl
DET2 & merger det.  & 1.20 & 0.62 & 1.55 & 13.7&1.27 & 0.35 & 180 \nl
DET2E6 & merger det.  & 1.80 & 0.62 & 1.33 & 19.8 &3.30 & 1.04 & 300 \nl
\enddata

\tablenotetext{a}{t$_{B}$ is the risetime to reach B band maximum.}
\tablenotetext{b}{\.{E}$_{\gamma}$ refers to the gamma energy deposition
rate, \.{E}$_{e+}$ refers to instantaneous deposition of the positron
kinetic energy.}

\end{deluxetable}

\begin{deluxetable}{llcclllc}
\footnotesize
\tablecaption{Summary of Observations. \label{sumobs}
\tablenotemark{b}}
\tablewidth{0pt}
\tablehead{
 & \colhead{Host} & \colhead{Photometric}  & \colhead{Distance} &
\colhead{E(B-V)} & \colhead{Suggested} & & \\
\colhead{SN} & \colhead{Galaxy} & \colhead{Bands} & 
\colhead{Spectra} &  \colhead{(Mpc)} & 
\colhead{(m)} & \colhead{Models} & \colhead{Bolometry} }
\startdata
92A & NGC 1380 & UBVRI & 2 & 13.5$\pm$0.7 (3) & 0.00 (1) & DD4 (2) & 1 \nl
    &          &  (1)  &   & 16.9 (4)         & 0.1  (2) & DD23C (6) & \nl
    &          &       &   & 20.9$\pm$3.6 (5) & 0.00 (31),(5)&       & \nl
91bg& NGC 4374 & UBVRI & 8,9&16.6 (10)        & 0.05 (9) & WD065 (11)&8 \nl
    &          & (7,8,9)&  & 21.1$\pm$3.6 (5) & 0.00 (5)& PDD5 (12) & \nl
    &          &       &   & 16.4$\pm$0.7 (31)& 0.00 (31)& ONeMg (13)& \nl
90N & NGC 4639 &UBVRI & 15&19.1$\pm$2.5 (31)& $\leq$0.15 (18)&W7DN (19)&1 \nl
  &      & (14)   &    &20.6 (16)& 0.01 (31)&DET2E2/4,PDD3 (20) \nl
  &      &        &    &25.5$\pm$2.5 (51) &         &              & \nl   
72E & NGC 5253 & UBV  & 22 & 4.1$\pm$0.1 (24) & 0.05 (25)& W7 (28) & 23 \nl
    &          & (21,22) & & 2.8$^{+0.7}_{-0.3}$ (25) & 0.02 (27) &M35 (20)
                                                                    &   \nl
91T & NGC 4527 & UBVRI &9,26&12-14 (25)      & 0.1-0.2 (25)&W7DT (33) & 1 \nl
    &          & (14,29,30)& &13.0 (32)      & 0.2 (9)     &PDD3, DET2E2 (20) 
                                                                     &   \nl
    &          &           & &               & 0.00 (31)   & SC1.1 (34)& \nl
    &          &           & &               & 0.13 (26)   & DD4 (35) &\nl
93L & IC 5270  & UBVRI &   & 20.1 (4)        & 0.2 (4)     & 
W7\tablenotemark{a} (4) & \nl
    &          &  (4)  &   &                 &             &      &    \nl
37C & IC 4182  & BV    & 37 & 4.9$\pm$0.2 (38)& 0.0$\pm$0.14 (36)& N32, W7 
                                                                        & \nl
    &          & (36)  &    & 2.5$\pm$0.3 (39)&0.13 (39)  & DET2 (20) &\nl
89B & NGC 3627 & UBVRI & 40 & 7.6 (31)       & 0.37 (40)   & M37, M36 (20) 
                                                                     &1 \nl
    &          &  (40) &    & 11.8$\pm$1.0 (41)&           &         & \nl
    &          &       &    & 11.6$\pm$1.5 (42)&           &        &   \nl
86G & NGC 5128 & UBVRI & 45 & 3 (46)  & 1.09$\pm$0.02 (25) & N32, W7 (20)&\nl
    &          & (43,44,45)& & 6.9 (24) & 0.9 (43)         &         &  \nl
    &          &       &    & 3.1$\pm$0.1 (47) & 0.88 (49) &     &    \nl
    &          &       &    & 3.6$\pm$0.2 (48) & 1.1 (45)  &    &     \nl
94D & NGC 4526 & UBVRI & 50 & 16.8 (4)      & 0.03 (18)    & M36 (20)&\nl
    &          &(50,17,30)& & 13.7 (50)     &              & SC0.9 (34)&\nl
    &          &       &    &               &              & W7 (4)  &  \nl

\enddata

\tablenotetext{a}{W7 was scaled to 1.0 M$_{\odot}$ to fit 93L.}
\tablenotetext{b}{REFERENCES.-(1) Suntzeff 1996; (2) Kirshner et al. 1993; 
(3) Tonry 1991; (4) Cappellaro et al. 1997; (5) Branch et al. 1997; 
(6) H\H{o}flich 1998; (7) Leibundgut et al. 1993; (8) Turatto et al. 1996; 
(9) Filippenko et al. 1992; (10) Tully 1988; (11) Ruiz-LaPuente et al. 1993; 
(12) H\H{o}flich 1996; (13) Nomoto et al. 1996; (14) Lira et al. 1998; 
(15) Jeffery et al. 1992; (16) Saha et al. 1997; (17) Tanvir et al. 1997;
(18) Hamuy et al. 1995; (19) Shigeyama et al. 1992; 
(20) H\H{o}flich \& Khokhlov 1996; (21) Ardeberg \& de Grood 1973; 
(22) Kirshner \& Oke 1975; (23) Axelrod 1980; (24) Sandage et al. 1994; 
(25) Ruiz-LaPuente \& Filippenko 1996; (26) Phillips et al. 1992; 
(27) Della Valle \& Panagia 1992; (28) Ruiz-LaPuente \& Spruit 1997; 
(29) Schmidt et al. 1994; (30) Cappellaro 1998; (31) Phillips 1993;
(32) Tully et al. 1992; (33) Yamaoka et al. 1992; (34) Liu et al. 1997; 
(35) Pinto \& Eastman 1996; (36) Schaefer 1994; (37) Minkowski 1939;
(38) Sandage et al. 1996; (39) Pierce et al. 1992; (40) Wells et al. 1994;
(41) Tammann et al. 1997; (42) Branch et al. 1996; (43) Phillips et al. 1987;
(44) Phillips et al. 1997; (45) Cristiani et al. 1992; 
(46) Hesser et al. 1984; (47) Tonry \& Schecter 1990; 
(48) Jacoby et al. 1988; (49) Rich 1987; (50) Patat et al. 1996;
(51) Saha et al. 1997.} 

\end{deluxetable}

\begin{deluxetable}{ll|cc|c}
\footnotesize
\tablecaption{Model Fits to SN 1992A. \label{table2}}
\tablewidth{0pt}
\tablehead{ \colhead{Model} & &
\multicolumn{2}{l}{Bolometric (55$^{d}$ -450$^{d}$)} & 
\colhead{V Band (55$^{d}$ -950$^{d}$)} \\
\colhead{Name} & \colhead{Type \tablenotemark{a}} &
\colhead{$\chi^{2}$/DOF} & \colhead{Probability} & \colhead{$\chi^{2}$/DOF} }
\startdata
W7 &    R1\%  & 0.84  & 0.68              & 23.9 \nl
   &    R3    & 2.86 & 2.4 x 10$^{-6}$    & 43.7 \nl
   &    T     & 2.98  & $<$10$^{-6}$      & 49.0 \nl
DD4 &   R1\%  & 0.86  & 0.66              & 26.4 \nl
   &    R3    & 1.56  & 0.04              & 34.5 \nl
   &    T     & 2.13  & 8.2 x 10$^{-6}$   & 43.1 \nl
M39 &   R1\%  & 0.98  & 0.48              & 22.1 \nl
   &    R3   & 1.02  & 0.43               & 23.4 \nl
   &    T     & 1.42  & 0.08              & 29.4 \nl
DD23C & R1\% & 0.73  & 0.84               & 23.9 \nl
   &    R3   & 1.33  & 0.13               & 30.9 \nl
   &    T    & 2.03  & 1.8 x 10$^{-3}$    & 39.9 \nl
W7DT &  R1\% & 1.28   & 0.16              & 30.1 \nl
   &    R3   & 1.92    & 3.8 x 10$^{-3}$  & 44.5 \nl
   &    T    & 6.05 & $<$10$^{-6}$        & 90.7 \nl
PDD3 &  R1\% & 1.14  & 0.28               & 37.4 \nl
  &     R3   & 3.66 & $<$10$^{-6}$        & 64.6 \nl
   &    T    & 5.32 & $<$10$^{-6}$        &  83.0 \nl
PDD54 & R1\% & 1.06   & 0.38              &   24.1 \nl
  &     R3   & 1.19     & 0.23            & 24.3 \nl
   &    T    & 1.73       & 0.01          &  26.5 \nl
PDD1b&  R1\% & 4.12 &$<$10$^{-6}$         & 61.2 \nl
  &     R3   & 4.09 &$<$10$^{-6}$         & 60.5 \nl
   &    T    & 4.15 &$<$10$^{-6}$         &  61.8  \nl
WD065 & R1\% & 3.34  &$<$10$^{-6}$        & 64.0  \nl
  &     R3   & 10.1 & $<$10$^{-6}$        &   130.9 \nl
   &    T    & 11.3 & $<$10$^{-6}$        &  153.2  \nl 
HED6 &  R1\% & 2.65 & 1.4 x 10$^{-5}$     & 53.9 \nl
  &     R3   & 10.0 & $<$10$^{-6}$        &  140.2 \nl
   &    T    & 14.5 & $<$10$^{-6}$        &  190.4 \nl 
HED8 &  R1\% & 2.00   & 2.1 x 10$^{-3}$   & 41.6 \nl
 &      R3   & 5.97   & $<$10$^{-6}$      & 83.4 \nl
 &      T    & 11.4    & $<$10$^{-6}$     & 151.1 \nl  
HECD &  R1\% & 1.16    & 0.26             & 33.5 \nl
 &      R3   & 2.62  & 1.7 x 10$^{-6}$    & 51.2 \nl
 &      T    & 5.79 & $<$10$^{-6}$        &  88.8 \nl
ONeMg & R1\% & 6.00  &   $<$10$^{-6}$     & 98.4 \nl
 &      R3   & 24.1 &    $<$10$^{-6}$     & 342.7 \nl
 &      T    & 33.8     & $<$10$^{-6}$    & 448.6 \nl
DET2   &R1\% &  1.91 &  4.0 x 10$^{-3}$   &   40.4 \nl
 &      R3   & 4.86  &   $<$10$^{-6}$     & 71.6 \nl
 &      T    & 8.7     &  $<$10$^{-6}$    &  118.7 \nl 
DET2E6 &R1\% & 1.27  &    0.16             & 19.7 \nl
 &      R3   &  1.21          & 0.22         & 19.4 \nl
 &      T    &  1.14         &    0.28       & 18.8 \nl
\enddata
\tablenotetext{a}{R1\% refers to 1\% ionized ejecta with a radial
 magnetic field, R3 refers to triply ionized ejecta with a radial
field, T refers to 1\% ionized ejecta with a trapping field.}
\end{deluxetable}

\begin{deluxetable}{lcccc}
\footnotesize
\tablecaption{Positron Yields for SN Ia Models. 
\tablenotemark{a}  \label{table3} }
\tablewidth{0pt}
\tablehead{
\colhead{Model} & \colhead{Radial} &\colhead{Radial} &
\colhead{Trapped} &\colhead{Trapped} \nl
& \colhead{Survival} & \colhead{Yield} & 
 \colhead{Survival} & \colhead{Yield} \nl
& \colhead{Fraction [\%]} & \colhead{[10$^{52}$ e+]} &
\colhead{Fraction [10$^{-4}$]} & \colhead{[10$^{50}$ e+]} 
}
\startdata
W7 & 1.8$-$5.5 & 4.4$-$13.1 & 0.2$-$2.8 & 0.4$-$6.8 \nl
DD4 & 1.5$-$4.6 & 3.8$-$11.6 & 0.04$-$1.1 & 0.1$-$2.8 \nl
M35 & 2.3$-$5.7 & 6.2$-$15.3 & 1.7$-$10.4 & 4.6$-$28.1 \nl
M36 & 2.0$-$5.7 & 4.9$-$13.7 & 1.0$-$17.6 & 2.5$-$42.8 \nl
M39 & 1.6$-$3.8 & 1.2$-$5.2 & 0.3$-$2.9 & 0.4$-$4.0 \nl
DD202C & 1.3$-$4.0 & 6.9$-$8.4 & 0.01$-$0.5 & 0.03$-$1.0 \nl
DD23C & 1.4$-$4.2 & 3.3$-$10.1 & 0.08$-$1.6 & 0.2$-$3.8 \nl
W7DN & 2.9$-$6.2 & 7.4$-$15.8 & 24.0$-$62.8 & 61$-$160 \nl
W7DT & 5.1$-$9.2 & 15.9$-$28.5 & 42.7$-$64.8 & 133$-$187 \nl
PDD3 & 1.6$-$5.0 & 3.8$-$12.0 & 0.07$-$1.7 & 0.18$-$4.1 \nl
PDD54 & 0.4$-$2.7 & 0.3$-$2.1 & 0.0$-$0.3 & 0.0$-$0.2 \nl
PDD1b & 0.0$-$0.1 & 0.0$-$0.1 & 0.0$-$0.0 & 0.0$-$0.0 \nl
HED6 & 2.4$-$8.6 & 2.4$-$8.6 & 2.2$-$52.6 & 2.3$-$53 \nl
HED8 & 3.6$-$9.1 & 6.6$-$16.9 & 6.3$-$58.2 & 11.8$-$109 \nl
HED9 & 0.9$-$7.6 & 2.2$-$19.5 & 11.6$-$86.8 & 30.0$-$223 \nl
HECD & 4.3$-$9.7 & 12.5$-$26.8 & 64.8$-$259 & 187$-$750 \nl
ONeMg & 3.5$-$11.3 & 2.2$-$7.2 & 0.5$-$15.6 & 0.3$-$10.1 \nl
DET2 & 2.7$-$6.3 & 6.9$-$16.2 & 1.1$-$9.8 & 3.0$-$25.1 \nl
DET2E6 & 0.1$-$1.0 & 0.3$-$2.4 & 0.0$-$0.0 & 0.0$-$0.1 \nl
\enddata
\tablenotetext{a}{Survival refers to escape and/or survival until
2000$^{d}$}
\end{deluxetable}

\clearpage

\clearpage

\clearpage

\clearpage

\clearpage

\clearpage

\begin{figure}
\plotone{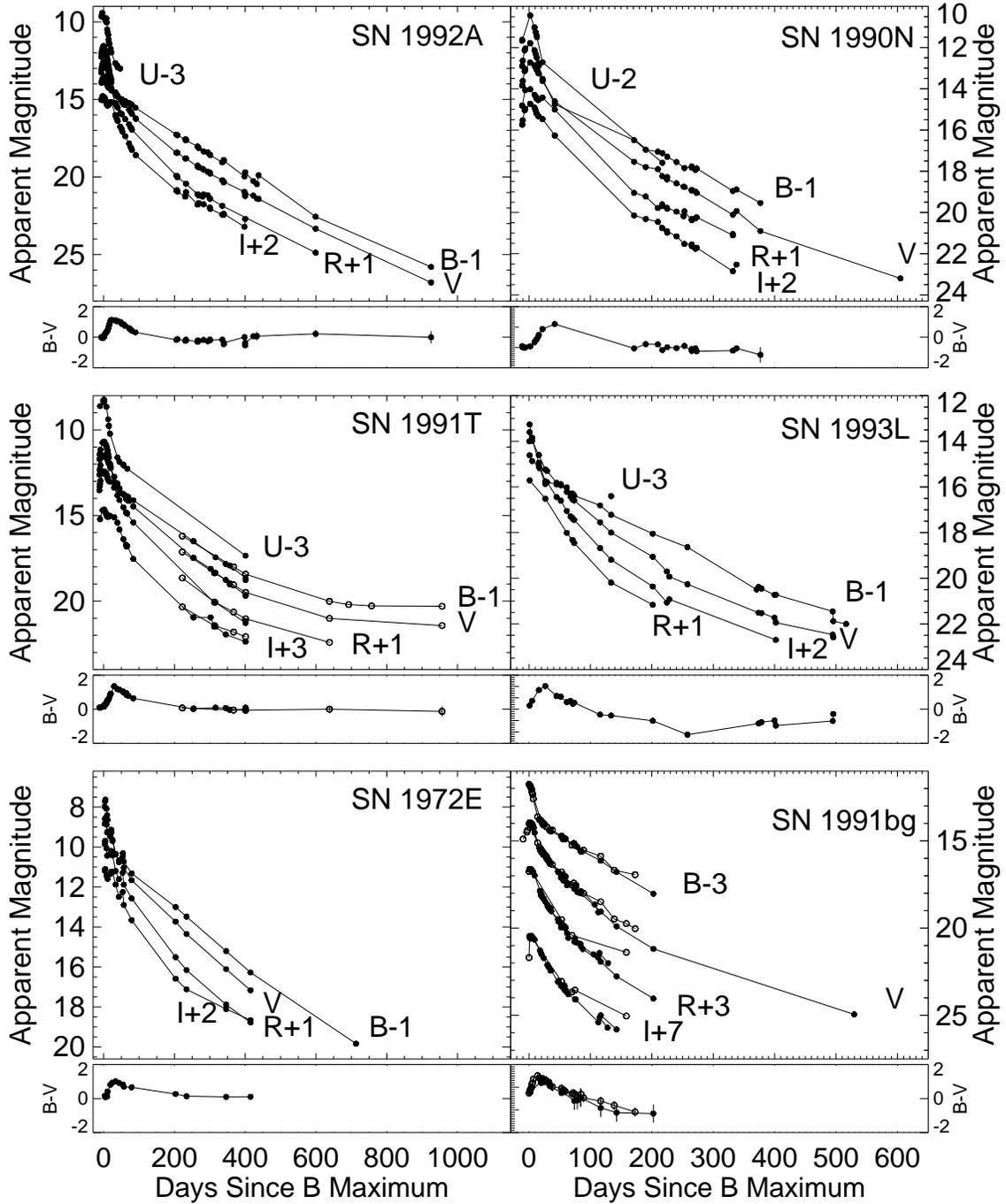}
\vspace{-1.0cm}
\caption{The multi-band light curves of the six SNe used in this study;
1992A (Suntzeff 1996), 1990N (Lira 1998), 
1972E (Kirshner \& Oke 1975; Ardeberg \& de Groot 1973; Jeffery 1998), 
 1991T (Lira 1998; Schmidt et al. 1994), 
1993L (CAPP), 1991bg (Leibundgut 
et al. 1993; Turratto et al. 1996; Filippenko
et al. 1992).
All show considerable color evolution before 100$^{d}$, all but
SN1991bg settle in to
constant values of the color indices at later times. \label{cev} }
\end{figure}

\begin{figure}
\plotone{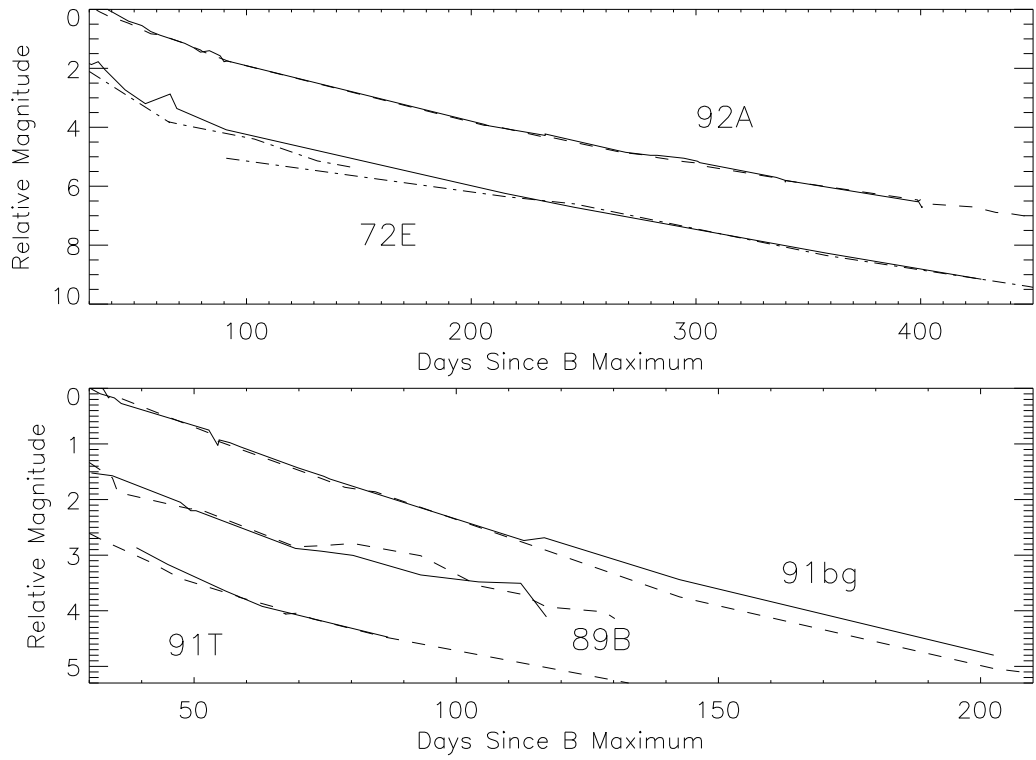}
\caption{The bolometric magnitudes (solid lines)
 of five SNe compared with the V (dashed lines) 
 and B (dot-dashed lines) band data. \label{bolvsv}}
\end{figure}

\begin{figure}
\plotone{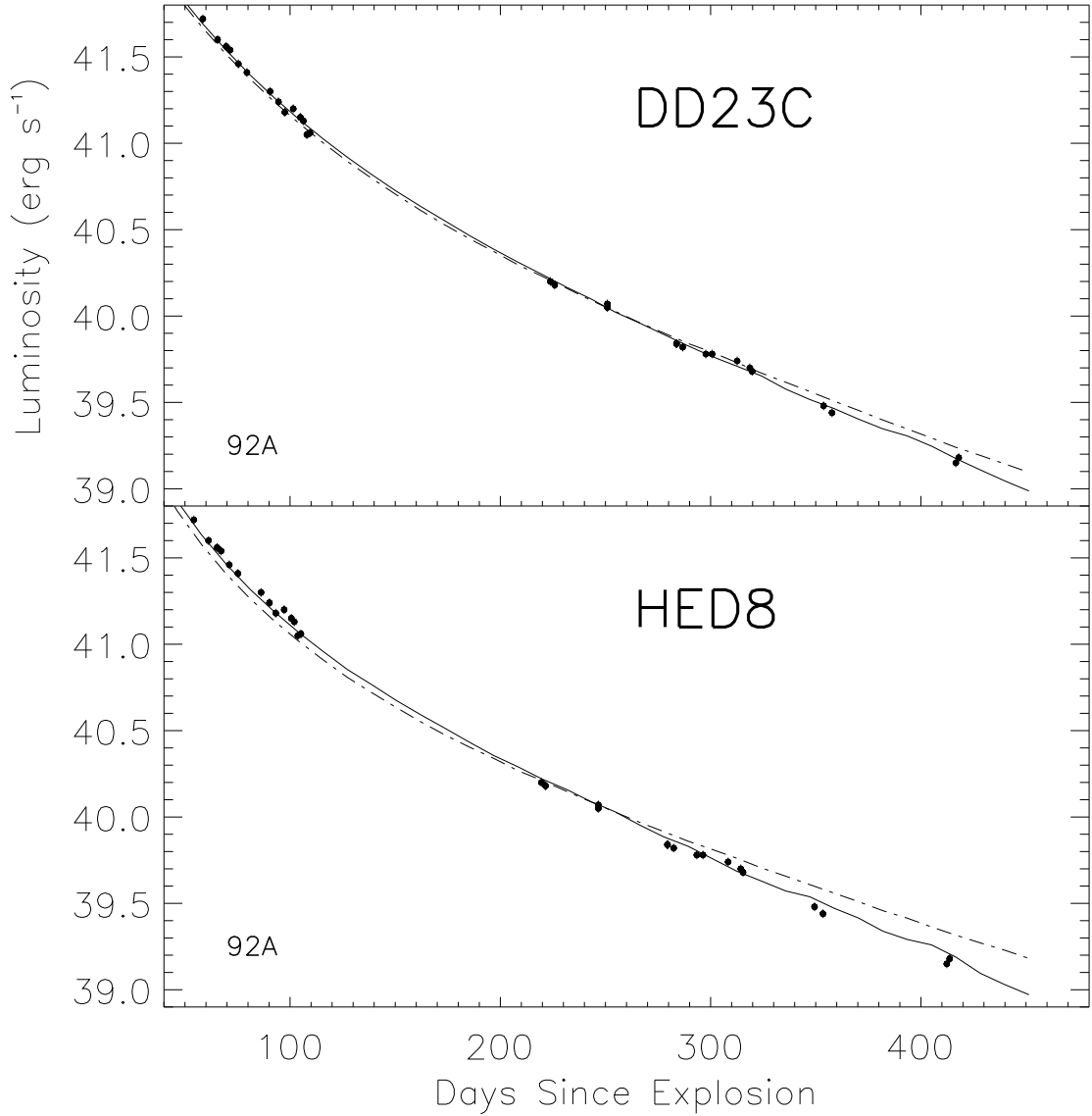}
\caption{The models DD23c and HED8 
 fit to the bolometric light curve of SN 1992A 
(Suntzeff 1996). The solid lines are model-generated 
bolometric light curves assuming a radial field geometry and 1\% 
ionization, 
the
 dot-dashed lines assume a positron trapping field. 
 The early data fixed
the normalization and demonstrates that without low-ionization 
positron escape, low-mass models remain too
bright to fit the 300$^{d}$ and later data. The model DD23c 
fits the data well for the radial field scenario. 
\label{bol92a} }
\end{figure}

\begin{figure}
\plotone{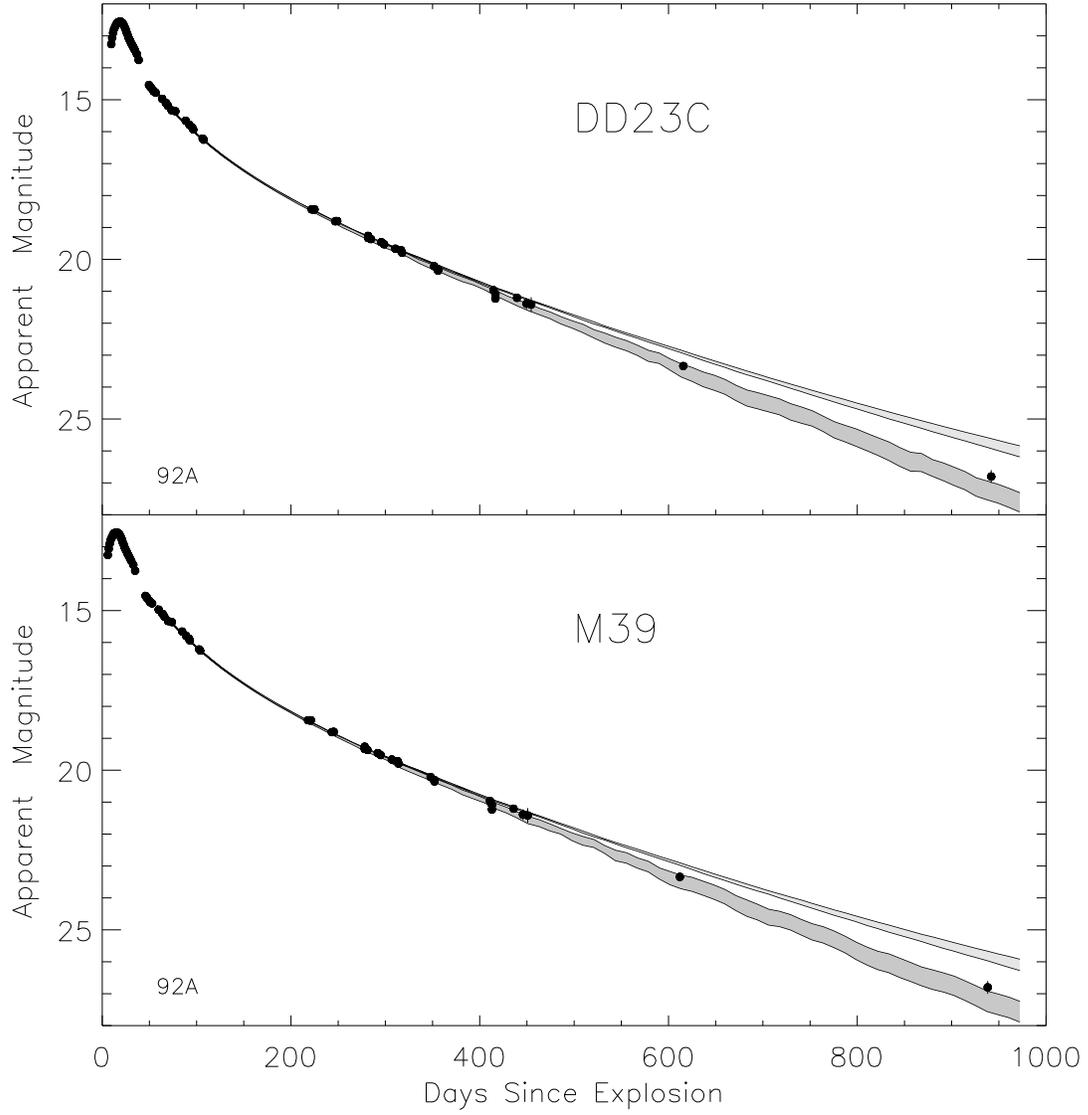}
\caption{The models DD23c and M39 fit to the V band data of SN 1992A
(Suntzeff 1996). The data
is fit much better by the radial scenario (dark shading)
 then by the trapping scenario (light shading).
\label{v92a} }
\end{figure}

\begin{figure}
\plotone{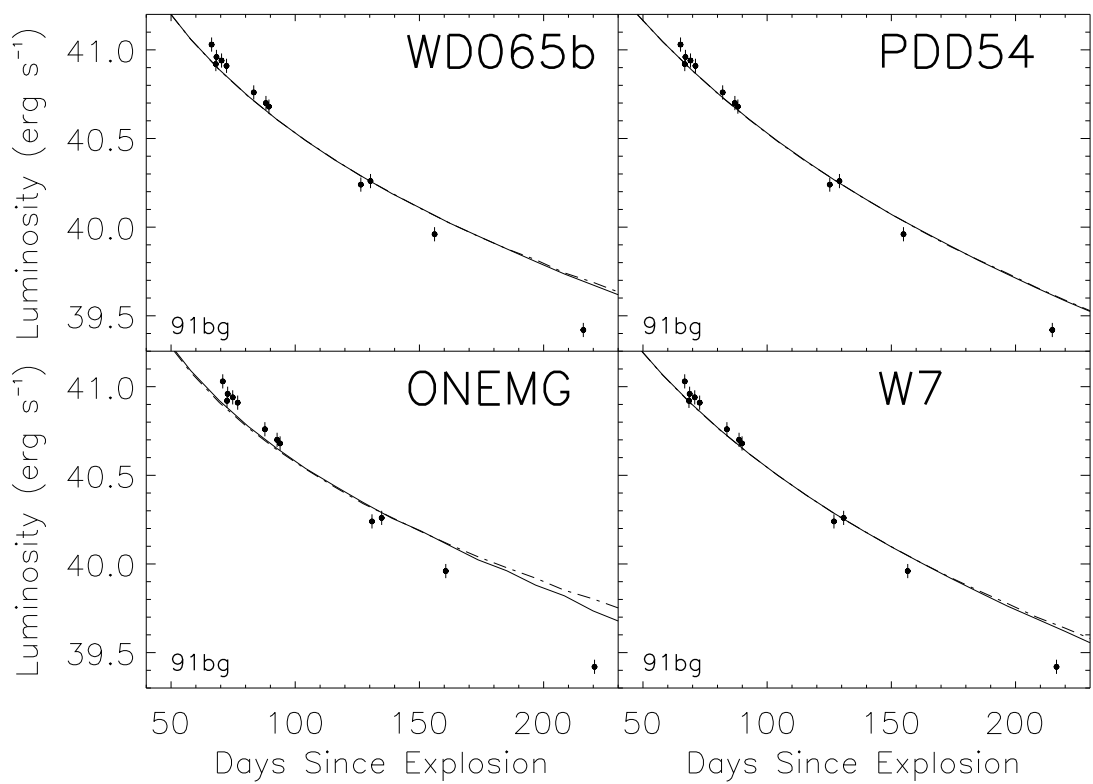}
\caption{The models WD065b, ONeMg, PDD54 and W7 fit to the
bolometric data for SN 1991bg (Turatto et al. 1996). The bolometric 
light curve was generated assuming E(B-V)=0.05$^{m}$.  
The model-generated bolometric light curve assuming a radial field 
geometry is shown as a solid line, the positron trapping light curve 
is shown as a dot-dashed line. 
 None of the models fit the entire
data-set, PDD54 and W7 approximate the early data.
\label{bol91bg}}
\end{figure}

\begin{figure}
\plotone{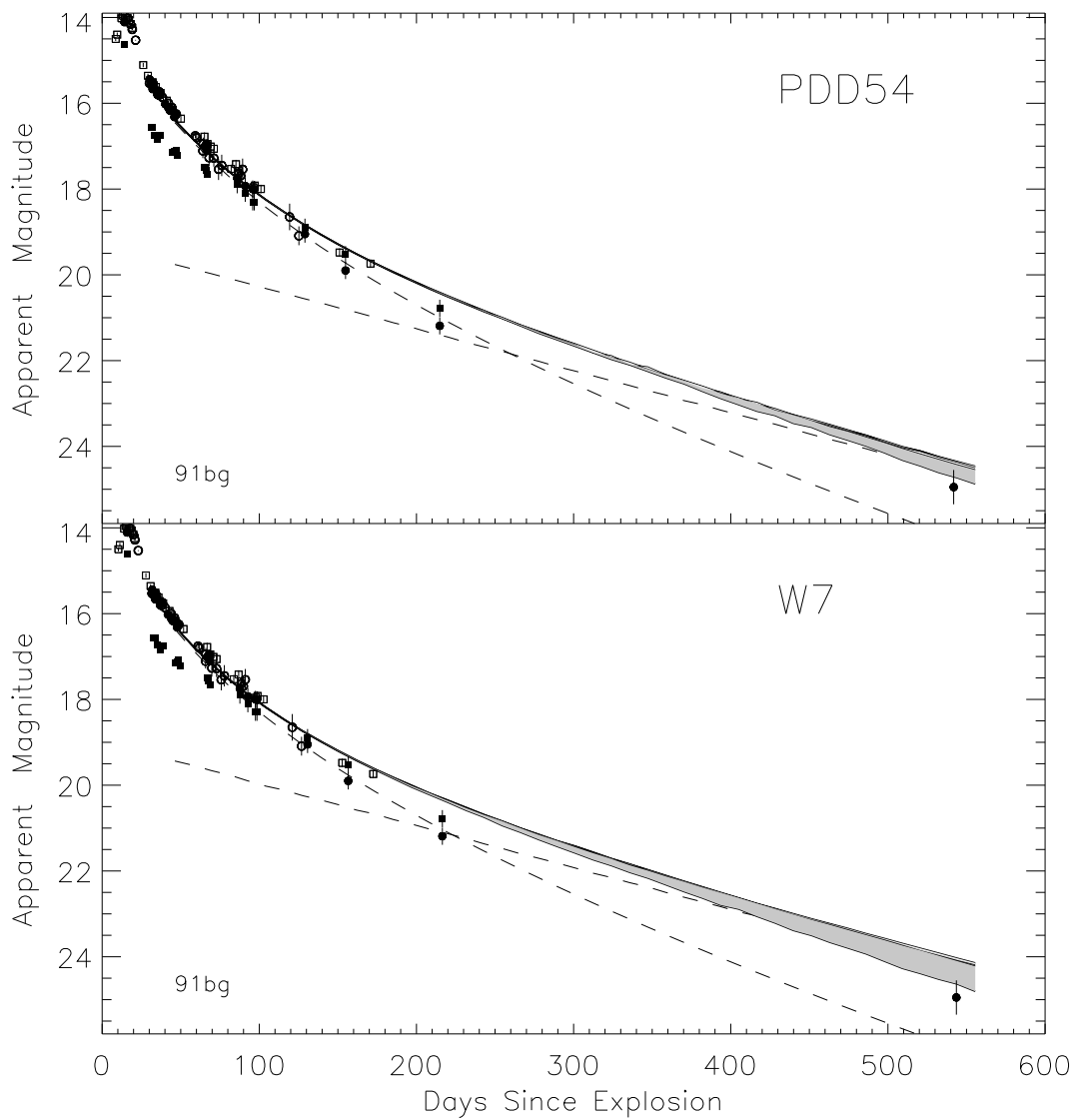}
\caption{The models PDD54 and W7 fit to the V band data for SN
1991bg (Turatto et al. 1996, Leibundgut et al. 1993, Filippenko et al.
1992).  The filled circles are the Turratto et al. data, the open
circles are the Fillippenko et al. data, the open boxes are the 
Leibundgut data. The filled boxes are the Turatto et al. B band data, 
with E(B-V)=0.25$^{m}$ (a value larger than claimed by Turatto). 
 The dashed lines
represent the individual energy deposition 
contributions from gamma photons and
positrons. Both models approximate the Leibundgut data from 
120$^{d}$ -170$^{d}$ and agree
with the 545$^{d}$ data point, but fit the Turatto data only if there is
no deposition of positron energy. 
\label{v91bg1}}
\end{figure}

\begin{figure}
\plotone{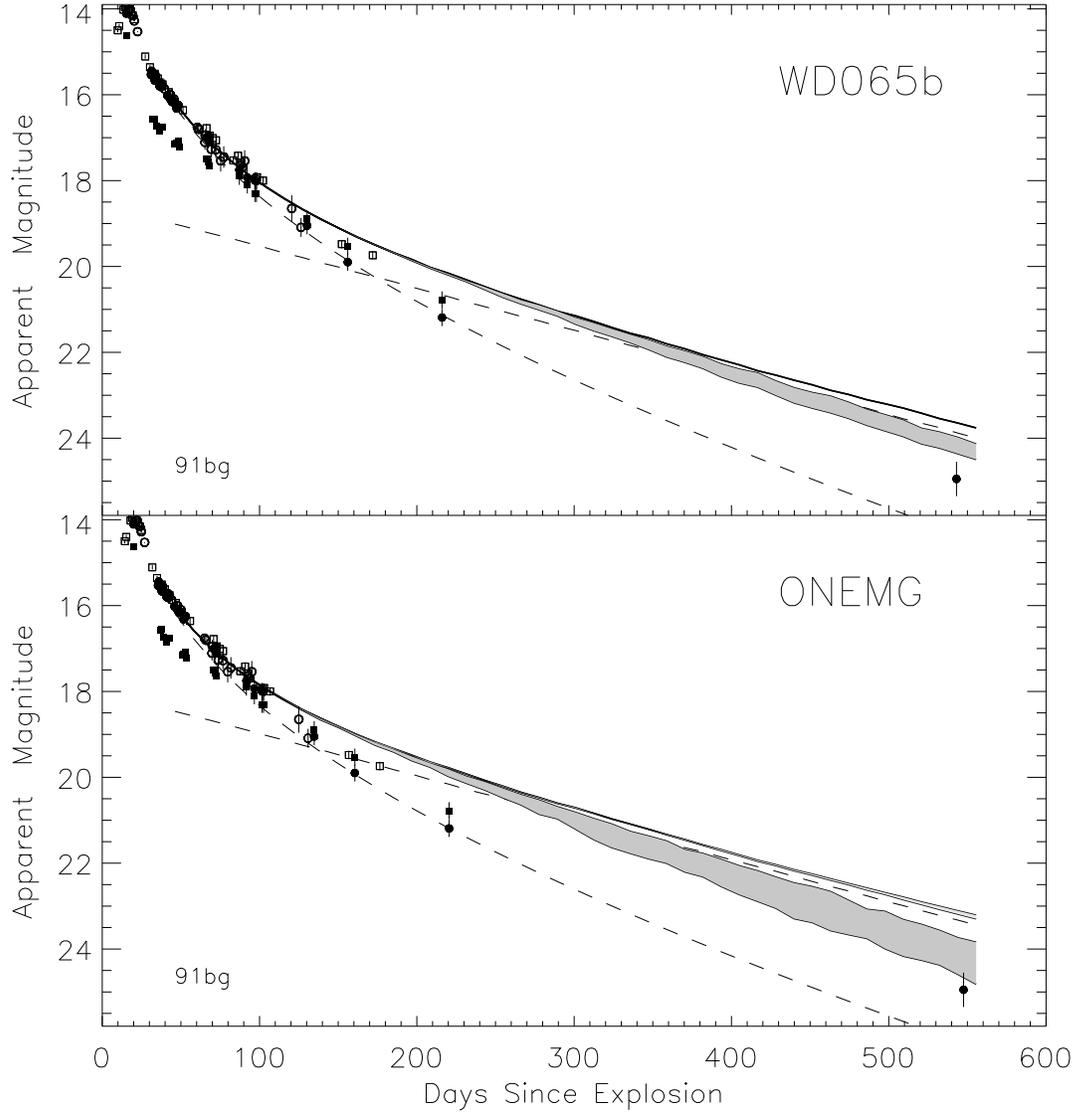}
\caption{The same as figure \protect\ref{v91bg1} for the models WD065b and
ONeMg. Both models remain too bright to fit any of the 130$^{d}$ data
and later. Both models can fit the Turatto data only if there is no
deposition of positron energy.
\label{v91bg2}}
\end{figure}

\begin{figure}
\plotone{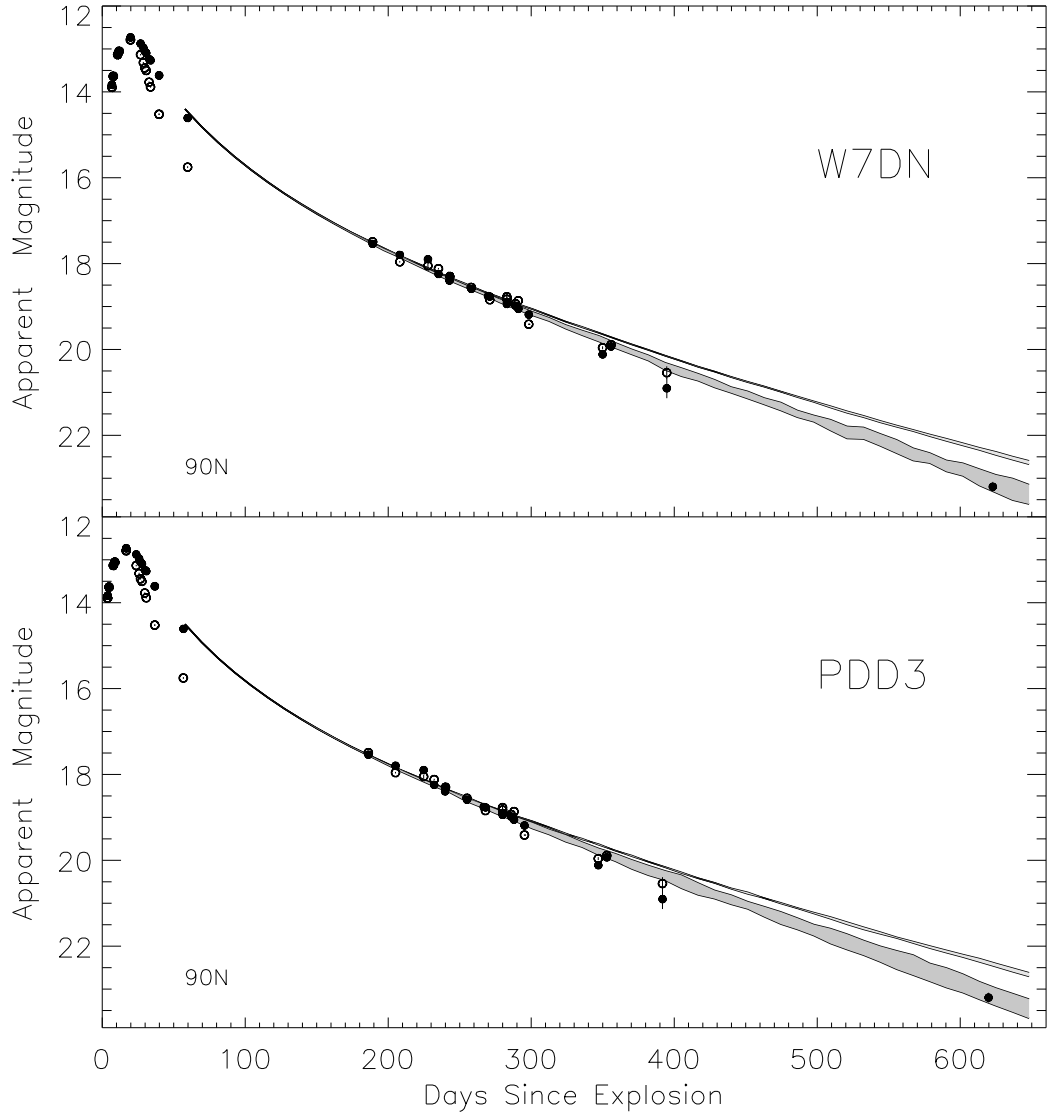}
\caption{The models W7DN and PDD3 fit to the B and V band data (open and
filled circles respectively) of SN 1990N
(Lira 1998). No extinction correction was applied to the B data. The
data shows appreciable scatter, but is better fit by the radial scenario
then by the trapping scenario. \label{bv90n} }
\end{figure}

\begin{figure}
\plotone{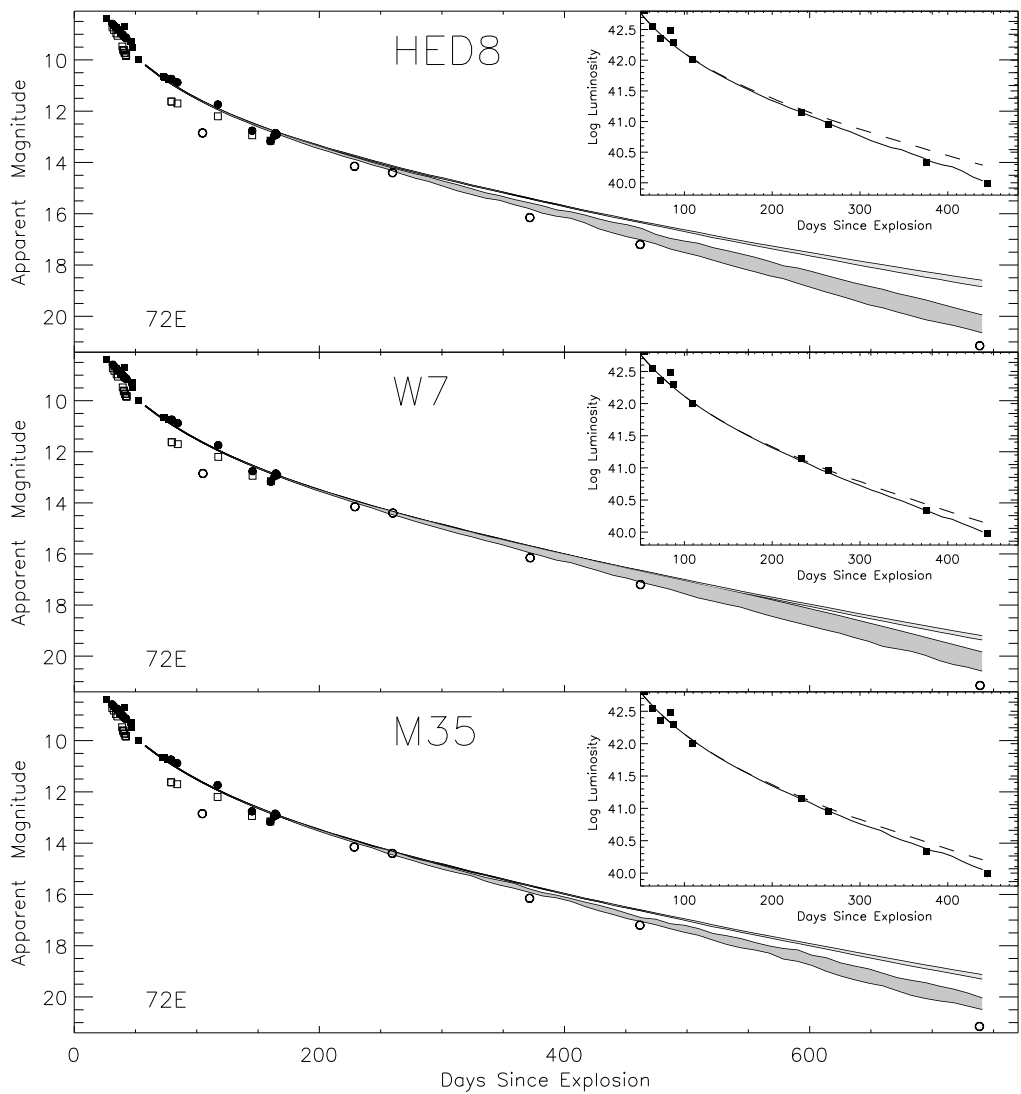}
\caption{The models HED8, W7 and M35 fit to B and V band data of SN 1972E.
The filled symbols are V data, the open symbols are B data. The
boxes are from Ardeberg \& de Groot (1973),
 the circles are from
Kirshner (1975). No extinction correction was applied to the B data. 
The inset is bolometric data
calculated by Axelrod (1980) from Kirshner's spectra, in units of erg
s$^{-1}$, the solid light curve assumes a radial field with 1\% 
ionization, the dashed light curve assumes a trapping field.
All three models show positron escape after
300$^{d}$.
\label{72e} }
\end{figure}

\begin{figure}
\plotone{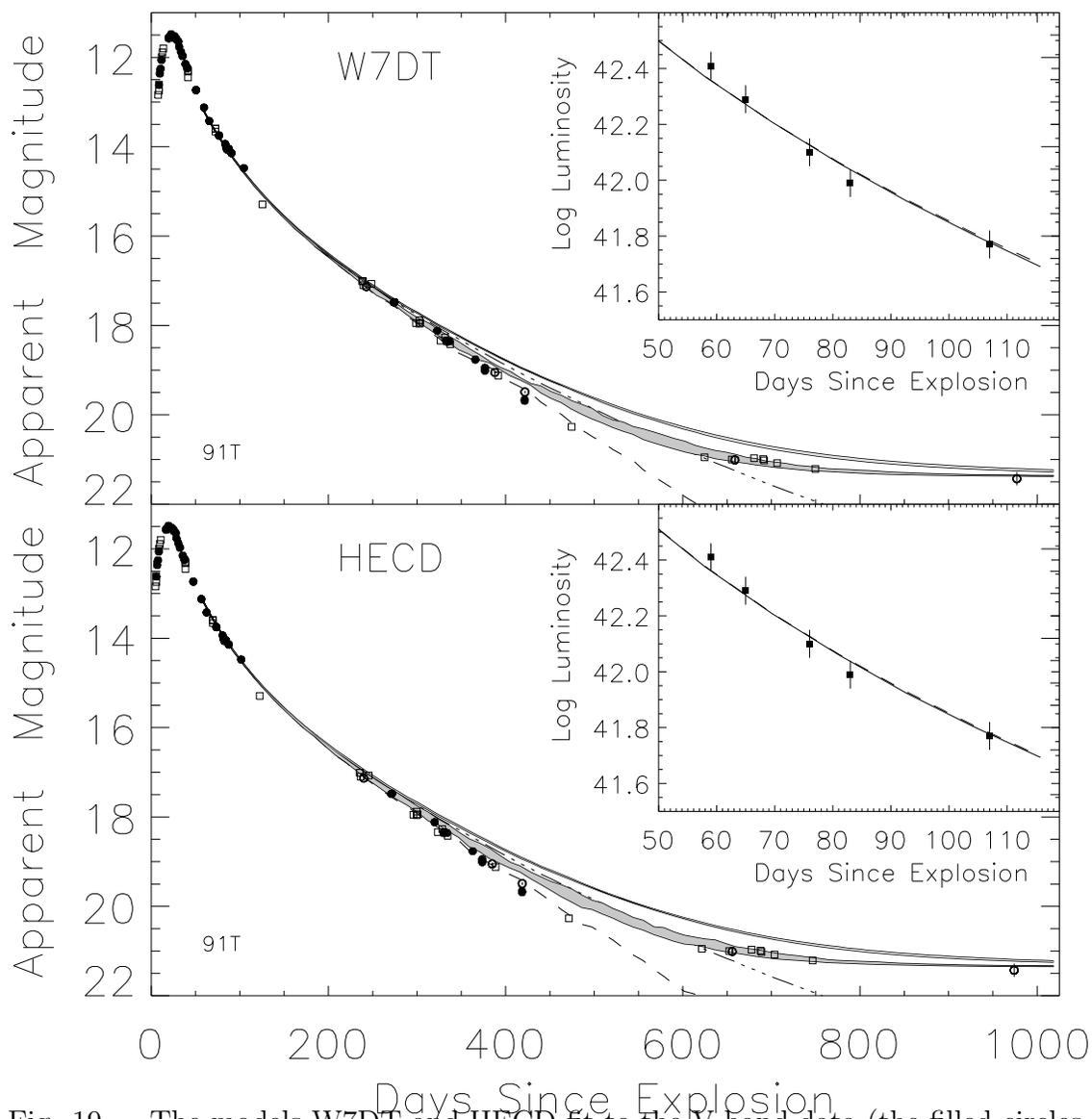}
\vspace{-1.0cm}
\caption{The models W7DT and HECD
 fit to the V band data (the  
filled circles are from Lira (1998), 
 the open circles are Schmidt et al. 1994, the open boxes are
Cappellaro (1997))
and bolometric light curves of SN 1991T (the bolometric light curve is
from Sunzteff 1996 and is in units of erg s$^{-1}$). 
 The light echo was treated to have a constant magnitude.
The dashed line shows the radial light curve with no light echo
contribution, dot-dashed line shows the 
same for the  trapping light curve. 
The trapping curves for both models remain too bright to fit the 
data, even with no contribution from a light echo.
 \label{91t} }
\end{figure}

\begin{figure}
\plotone{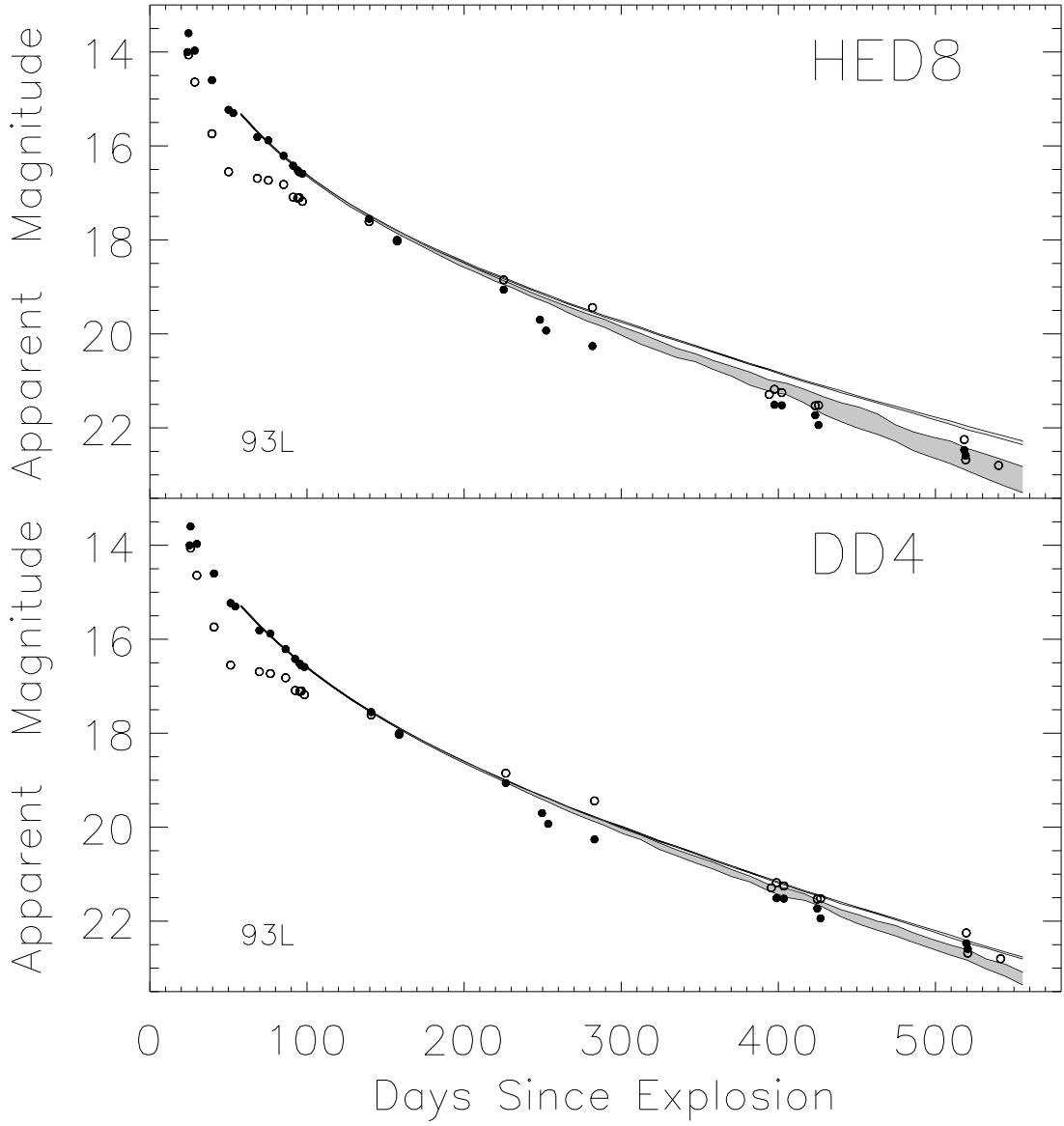}
\caption{The models HED8 and DD4 
fit to the B band (open circles) and V band (filled circles) 
light curves of SN 1993L (CAPP).
An extinction correction of 0.2$^{m}$ was applied to the B data. 
 The scatter is considerable, but for HED8, the 350$^{d}$+ data is
better fit by the radial scenario. For DD4, the results are
the same, but with much less separation between scenarios. \label{bv93l} }
\end{figure}

\begin{figure}
\plotone{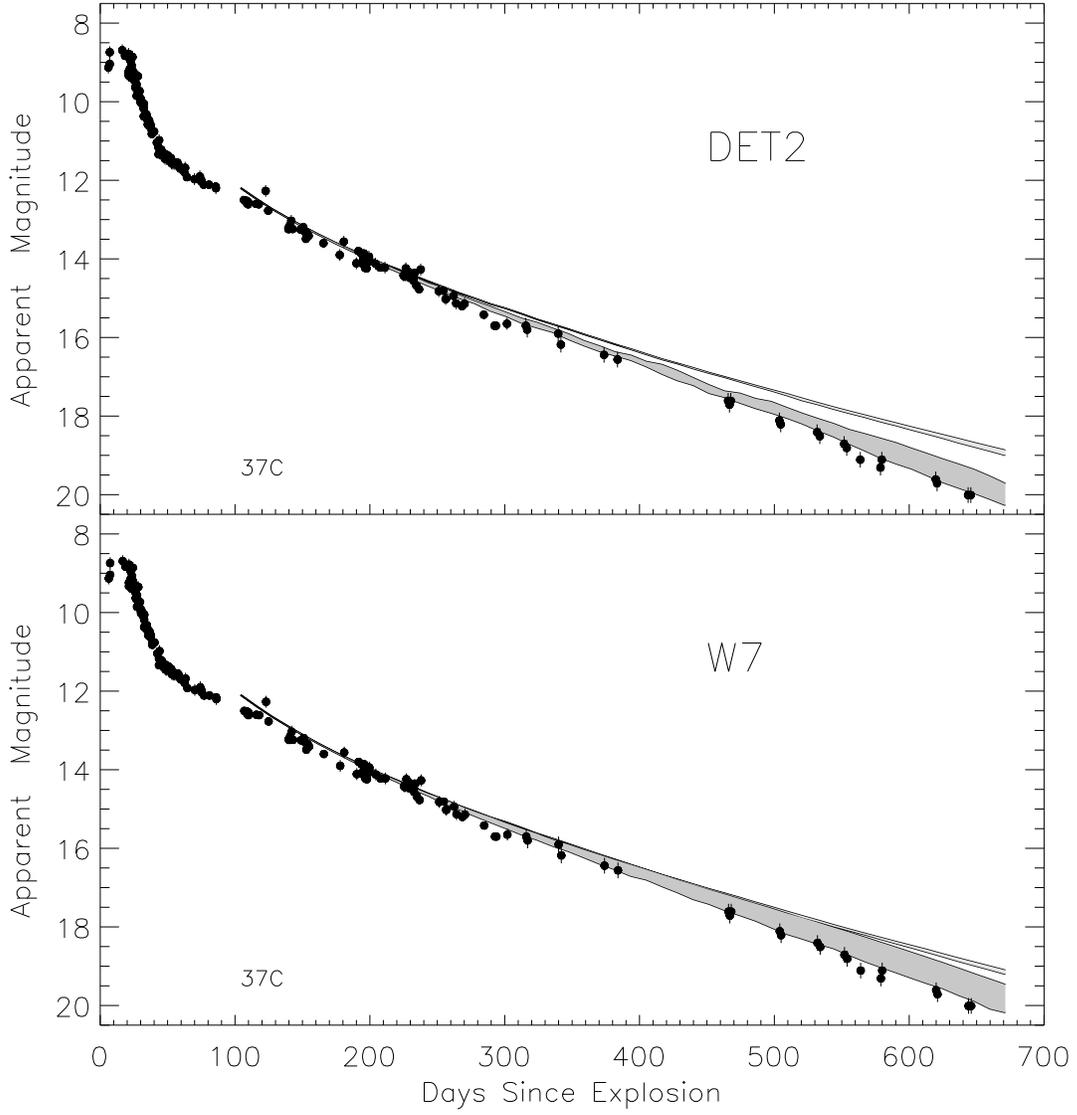}
\caption{The models DET2 and W7 fit to the B band light curve of SN
1937C (Shaefer 1994). There is evidence of positron escape for both
models. \label{b37c}}
\end{figure}

\begin{figure}
\plotone{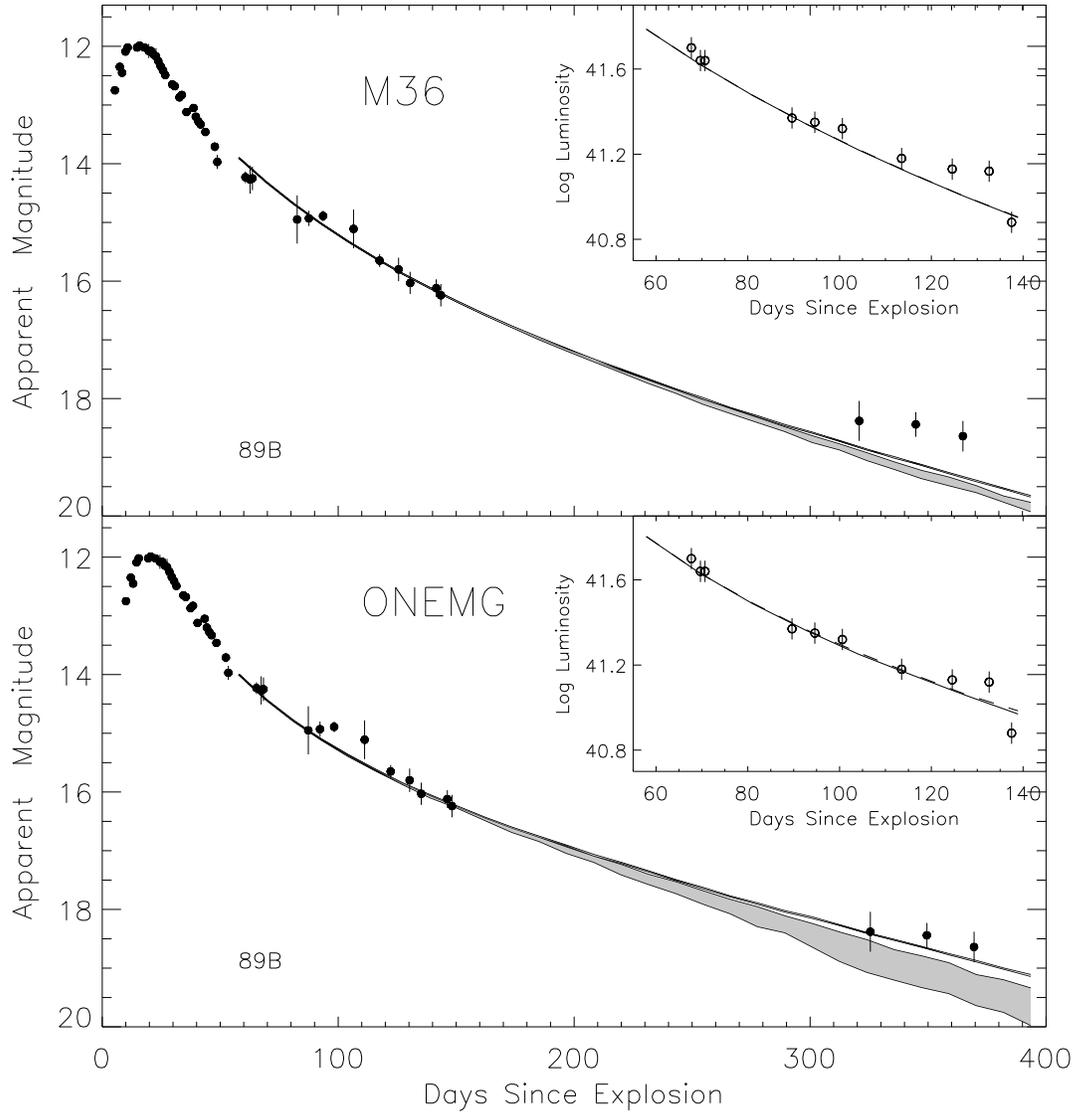}
\caption{The models M36 and ONeMg fit to the V band 
data of SN 1989B (Wells et al. 1994). The inset is bolometric data
calculated by Suntzeff (1996) (in units of erg
s$^{-1}$). The 60$^{d}$ -120$^{d}$
behavior was too erratic to be fit by modelling, but the SN settled
in to a smoother evolution after 120$^{d}$. M36 is clearly too faint
to explain the late data, the trapping curve of ONeMg comes close to
explaining the 320$^{d}$ -370$^{d}$ data. \label{89b}}
\end{figure}

\clearpage
\begin{figure}
\plotone{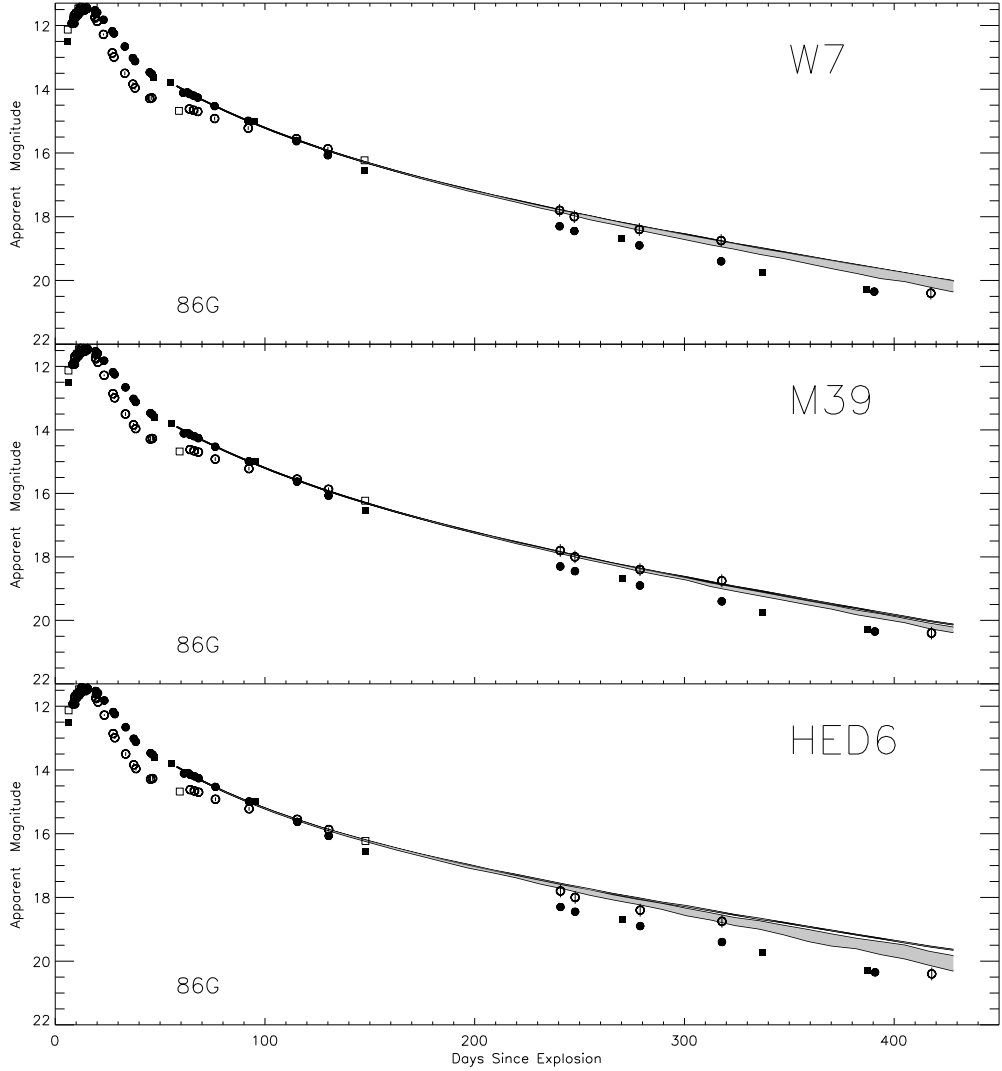}
\caption{The models W7, M39, HED6 fit to the B (open symbols)
  and V (filled symbols)  band 
data of SN 1986G. The circles are from Phillips et al. (1987), the 
squares are from Cristiani (1992). 
 When an extinction correction of 1.1$^{m}$ was applied to 
the B data, 
 W7 and M39 fit the data within the uncertainties, while 
HED6 remains too bright. The data does not extend to late
enough epochs to determine the positron transport,
but the escape scenario adequately explains the existing data. 
\label{bv86g}}
\end{figure}

\clearpage
\begin{figure}
\plotone{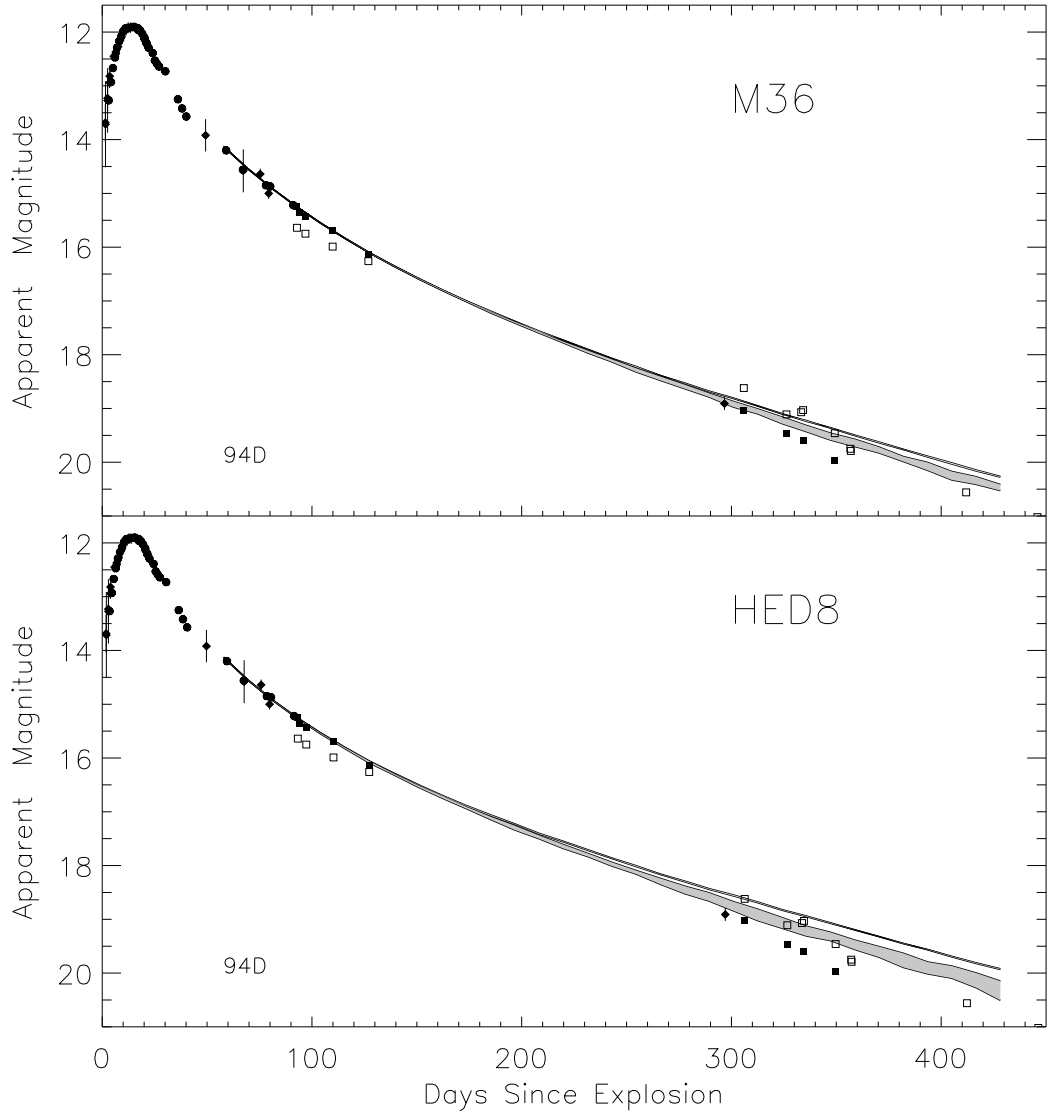}
\caption{The models M36 and HED8 fit to the B and 
V band data of SN 1994D. The early V data (filled circles) are from
Patat et al. (1996), the middle V data (filled diamonds) are Tanvir
(1997), the late V and B data (filled and open boxes) are CAPP and
Cappellaro (1998). No extinction correction was applied to the B data. 
The late data are fit by both models, 
differentiation between model types and positron transport scenarios
is not possible. \label{bv94d}}
\end{figure}

\clearpage
\begin{figure}
\plotone{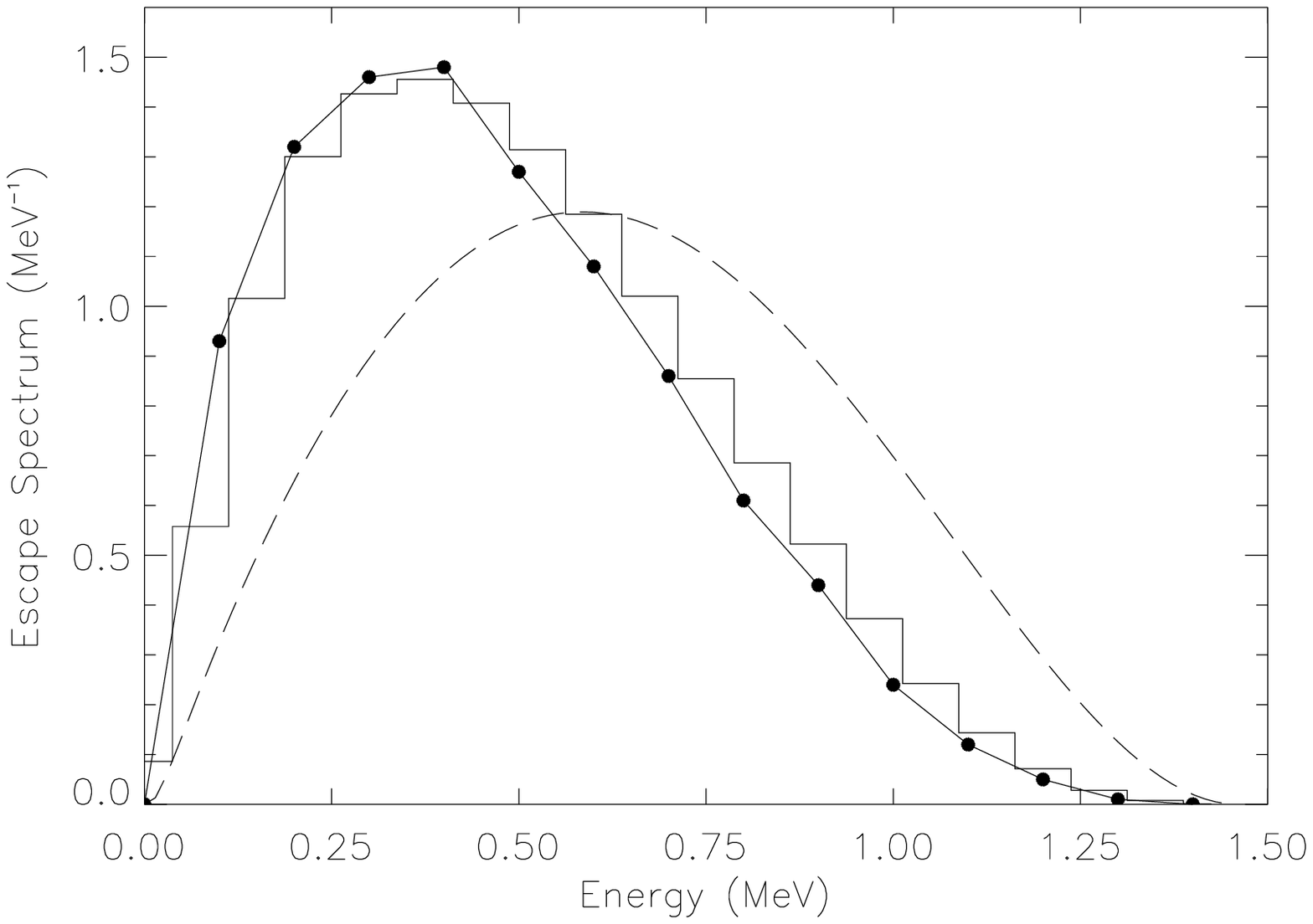}
\caption{The distribution of emitted positron kinetic energies as
estimated by Segre (1977, dotted line) compared to the
spectrum of escaping positrons from W7, as estimated by Chan and
Lingenfelter (filled circles, CL)
 and this work (solid histogram). Our
results agree with CL and demonstrate that the slowing of the positrons
leads to the mean energy shifting 
 from 632 keV to 494 keV. \label{emit}}
\end{figure}


\begin{thebibliography}{}
\bibitem{ahma97} Ahmad, I., Fischer, S.M., Bonino, G., Castagnoli, G.C.,
Kutschera, W., \& Paul, M. 1997, \prl, 80, 2550
\bibitem{arde73} Ardeberg A.L., \& de Grood, M.J. 1973, A \& A,28,295
\bibitem{arne69} Arnett, W.D. 1969, \apj, 157, 1369
\bibitem{arne80} Arnett, W.D. 1980, \apj, 237, 541 
\bibitem{arne82} Arnett, W.D. 1982, \apj, 253, 785 
\bibitem{axel80} Axelrod, T.S. 1980, Ph.D. thesis, University of
California at Santa Cruz
\bibitem{baad38} Baade, W. 1938, \apj, 88, 285
\bibitem{bazw38} Baade, W., \& Zwicky, F. 1938, \apj, 88, 411
\bibitem{boff98} Boffi, F.R. et al. 1998, \baas, 192, 6.07
\bibitem{bowe97}
Bowers, E.J.C., Meikle, W.P.S., Geballe, T.R., Walton,
N.A., Pinto, P.A., Dhillon, V.S., Howell, S.B., \& Harrop-Allin, M.K.
1997, LANL, 9707119
\bibitem{bran96} Branch, D., Romanishin, W., \& Baron, E. 1996,\apj,465,73
\bibitem{bran97} Branch, D., Nugent, P., \& Fisher, A. 1997 in 
Thermonuclear Supernovae, ed. P. Ruiz Lapuente, R. Canal, \& J. Isern
(Dordrecht: Academic Publishers), 715
\bibitem{burr90} Burrows, A., \& The, L.-S. 1990, \apj, 360, 626
\bibitem{capp97} 
Capellaro, E., Mazzali, P., Benetti, S., Danziger
 I.J., Turatto, M., Della Valle, M., \& Patat, F. 1997, \astap,328, 203 
\bibitem{capp97b} Capellaro, E., Turatto, M., Tsvetkov, D.Yu., 
Bartunov, O.S., Pollas, C., Evans, R., \& Hamuy, M. 1997, A\&A, 322, 431 
\bibitem{capp98} Cappellaro, E., et al. 1998, unpublished
\bibitem{chan93} Chan, K.-W., \& Lingenfelter, R.E. 1993, \apj, 405, 614 
\bibitem{colg70}Colgate, S. 1970, \apss, 8, 457 
\bibitem{colg80} 
Colgate, S., Petschek, A.G., \& Kreise, J.T. 1980, \apj, 237, L81
\bibitem{colg96}Colgate, S., Fryer, C.L., \& Hand, K.P. 1996,
in Thermonuclear Supernovae, ed. P. Ruiz Lapuente, R. Canal, \& J. Isern
(Dordrecht: Academic Publishers), 273
\bibitem{cris92} Cristiani, S. et al. 1992, \aap, 259, 63
\bibitem{dell92} Della Valle, M. \& Panagia, N. 1992,\aj,104,696
\bibitem{fili92} Filippenko, A.V., et al. 1992, \aj, 104, 1543
\bibitem{fili92b} Filippenko, A.V., et al. 1992, \apj, 384,L15
\bibitem{fran96} Fransson, C., Houck, J., \& Kozma, C. 1996, in 
Supernovae and Supernova Remnants, IAU
       Colloquium 145, eds.  R. McCray \& Z. Wang, p. 211
\bibitem{hamu95} Hamuy, M., Phillips, M.M., Maza., J., Suntzeff, N.B.,
Schommer, R.A., \& Avil\'{e}s, R. 1995, \aj, 109,1
\bibitem{hamu98} Hamuy, M., Pinto, P.A. 1998, LANL, 9812084
\bibitem{hess84} Hesser, J.E., Harris, H.C., van den Bergh, S., 
\& Harris, G.L.H. 1984, \apj, 276, 491
\bibitem{hofl93} 
H\H{o}flich, P., M\H{u}ller, E., \& Khokhlov, A. 1993, A \& A, 268, 570 
\bibitem{hofl96b} H\H{o}flich et al.  1996, \apj,472, L81  
\bibitem{hofl96}
H\H{o}flich, P., \& Khokhlov, A. 1996, \apj, 457, 500
\bibitem{hofl96c} H\H{o}flich, P. 1996, in 
Supernovae and Supernova Remnants, IAU
       Colloquium 145, eds.  R. McCray \& Z. Wang, p. 29
\bibitem{hofl95} H\H{o}flich, P. 1995, \apj, 443, 89
\bibitem{holf98} 
H\H{o}flich, P., Wheeler, J. C.,  \& Thielemann, F.-K. 1998, \apj, 495, 617 
\bibitem{hofl98b} H\H{o}flich, P., Private Communication
\bibitem{hoyl60} Hoyle, F. \& Fowler, W. A. 1960, \apj, 132, 565 
\bibitem{icru84} Stopping Powers for Electrons and Positrons, 
International Commission on Radiation Units and Measurements
 Report; 37 1984  
\bibitem{jaco88} Jacoby, G.H., Ciardullo, R., \& Ford, H.C. 1988, 
in The Extragalactic Distance Scale, BYU Press
\bibitem{jeff92} Jeffery, D.J., Leibundgut, B., Kirshner, R.P., 
Benetti, S., Branch, D., \& Sonneborn, G. 1992, \apj, 397, 304
\bibitem{jeff98} Jeffery, D.J. 1998, private communication
\bibitem{kirs75} Kirshner, R.P., \& Oke, J.B. 1975, \apj, 200, 574
\bibitem{kirs93} Kirshner, R.P., et al. 1993, \apj, 415, 589 
\bibitem{kuma97} Kumagai, S. 1997, private communication
\bibitem{leib91} Leibundgut, B., Kirshner, R.P., Filippenko, A.V., 
Shields, J.C., Foltz, C.B., Phillips, M.M., \& Sonneborn, G. 1991, \apj, 
371, L23
\bibitem{leib92} Leibundgut, B. \& Pinto, P.A. 1992, \apj, 401, 49
\bibitem{leib93} Leibundgut, B., et al. 1993, \aj, 105,301
\bibitem{leis95} 
Leising, M.D., Johnson, W.N., Kurfess, J.D., Clayton,
D.D., Grabelsky, D.A., Jung, G.V., Kinzer, R.L., Purcell, W.R.,
Strickman, M.S., The L.-S., \& Ulmer, M.P. 1995, \apj, 450, 805
\bibitem{lira95} Lira, P. Master's Thesis, University of Chile
\bibitem{lira98} Lira, P., et al. 1998,\aj,115,234
\bibitem{liu97} Liu, W., Jeffery, D.J., \& Schultz, D.R. 1997, \apj,483,L107 
\bibitem{liu97b} Liu, W., Jeffery, D.J., \& Schultz, D.R. 1997, \apj,486,L35
\bibitem{liu98} Liu, W., Jeffery, D.J., \& Schultz, D.R. 1998, \apj,494,812
\bibitem{mazz96} Mazzali, P.A., et al. 1996, \mnras, 284, 151
\bibitem{miln98} Milne, P.A. 1998, PhD thesis, Clemson University
\bibitem{miln99} Milne, P.A., Kurfess, J.D., \& Leising, M.D. 1999, in 
preparation
\bibitem{mink39} Minkowski, R. 1939, \apj, 89, 143
\bibitem{nomo84} 
Nomoto, K., Thielemann, F.-K., \& Yokoi, K. 1984, \apj, 286, 644
\bibitem{nomo96} 
Nomoto K. et al. 1996, in Supernovae and Supernova Remnants, IAU
       Colloquium 145, eds.  R. McCray \& Z. Wang, p. 49
\bibitem{pata96} Patat, F., Benetti, S., Cappellaro, E.,
Danziger, I.J., Della Valle, M., Mazzali, P.A., \& Turatto, M. 1996,
\mnras,278,111
\bibitem{phil87} Phillips, M.M., et al. 1987,\pasp, 99, 592
\bibitem{phil92} Phillips, M.M., Jacoby, G.H., Walker, A.R., Tonry,
J.L. \& Ciardullo, R. 1992, \baas, 24,749
\bibitem{phil97} Phillips, M.M. 1997, unpublished
\bibitem{phil92b} Phillips, M.M., Wells, L.A., Suntzeff, N.B., 
Hamuy, M., Leibundgut, B., Kirshner, R.P., \& Foltz, C.B. 1992,\aj,103,1632
\bibitem{phil93} Phillips, M.M. 1993, \apj, 413, L105
\bibitem{pier92} Pierce, M.J., Ressler, M.E., \& Shure, M.S. 1992, \apj, 
390, L45
\bibitem{pier95} Pierce, M.J., \& Jacoby, G.H. 1995, \aj, 110, 2885
\bibitem{pint96} Pinto, P.A., \& Eastman, R. 1996, LANL Preprints, 9611195 
\bibitem{purc97} Purcell, W.R., et al. 1997, \apj, 491, 725
\bibitem{rich87} Rich, R.M. 1987, \aj, 94, 651
\bibitem{royr68} Roy, R.R., Reed, R.D. 1968, Interactions of Photons
and Leptons with Matter, (New York: Academic Press)
\bibitem{ruiz92} Ruiz-Lapuente, \& P., Lucy, L.B. 1992,\apj,400,127
\bibitem{ruiz93} Ruiz-Lapuente, P., et al. 1993, \nat, 365, 728
\bibitem{ruiz96} Ruiz-Lapuente, P., \& Filippenko, A.V. 1996 in 
Supernovae and Supernova Remnants, IAU
       Colloquium 145, eds.  R. McCray \& Z. Wang, p. 33 
\bibitem{ruiz97} Ruiz-Lapuente, P., \& Spruit, H. 1997,\apj,500,360
\bibitem{saha97} Saha, A., Sandage, A., Labhardt, L., Tammann, G.A., 
Macchetto, F.D., \& Panagia, N. 1997,\apj,486,1
\bibitem{sand92} Sandage, A., Saha, A., Tammann, G.A.,
Panagia, N. \& Macchetto, F.D. 1992, \apj,401, L7
\bibitem{sand94} Sandage, A., Saha, A., Tammann, G.A., Labhardt, L.,
Schwengeler, H., Panagia, N. \& Macchetto, F.D. 1994, \apj, 423,L13
\bibitem{sand96} Sandage, A., Carlson, G., Kristian, J., Saha, A., 
\& Labhardt, L. 1996, \aj, 111, 1872
\bibitem{scha94} Schaefer, B.E. 1994, \apj,426,493
\bibitem{schm94} Schmidt,B.P., Kirshner,R.P., Leibundgut, B.,
Wells, L.A., Porter, A.C., Ruiz-Lapuente, P., Challis,
\& P., Filippenko, A.V. 1994, \apj, 434, L19 
\bibitem{segr77} Segre, E. 1977, Nuclei and Particles, 
(New York: Benjamin/Cummings)
\bibitem{shig92} Shigeyama, T., Nomoto, K., Yamaoka, H., \& Thielemann,
F.-K. 1992, \apj,386,L13
\bibitem{sunt96} Suntzeff, N.B., 1996, in Supernovae and Supernova
Remnants, ed. R. McCray, Z. Wang (Cambridge:Cambridge Press), p. 41
\bibitem{sunt98} Suntzeff, N.B., Phillips, M.M., et al. 1998, LANL, 
9811205
\bibitem{tamm97} Tammann, G.A., Sandage, A., Saha, A., Labhardt, L,
Macchetto, F.D., \& Panagia, N. 1997, in 
Thermonuclear Supernovae, ed. P. Ruiz Lapuente, R. Canal, J. Isern
(Dordrecht: Academic Publishers), 735
\bibitem{tanv97} Tanvir, N., Shanks, T., Dhillon, V., Lucy, J.,
Morris, P., Knapen, J., Balcells, M., Irwin, M., Casteneda, H.,
\& Walton, N. 1997, unpublished
\bibitem{the94} The, L.-S., Bridgeman, W. \& Clayton, D.D. 1994, \apjs, 
93, 531
\bibitem{tbb90} The, L.-S., Burrows, A., \& Bussard, R. 1990, ApJ, 352, 731
\bibitem{tonr91} Tonry, J.L. 1991, \apj, 373, L1
\bibitem{tonr90} Tonry, J.L., \& Schechter, P.L. 1990, \aj, 100, 1794
\bibitem{tull88} Tully, R.B. 1988, Nearby Galaxy Catalog, Cambridge
Univ. Press, Cambridge
\bibitem{tull92} Tully, R.B., Shaya, E.J., \& Pierce, M.J. 1992, \apjs, 
80, 479
\bibitem{tura96} 
Turatto, M.,Benetti, I.J., Capellaro, E., Danziger, I.J., Della
Valle, M., Gouiffes, C.,  Mazzali, P., \& Patat, F. 1996, \mnras, 283, 1 
\bibitem{well94} Wells, L.A., et al. 1994,\aj,106,2233
\bibitem{woos86} Woosley, S. E. \& Weaver, T. A., 1986, ARA\&A, 24, 205
\bibitem{yama92} Yamaoka, H., Nomoto, K., Shigeyama, T.,  \&
Thielemann, F.-K. 1992,  \apj, 393, L55 

\end{thebibliography}
\end{document}